\documentclass[preprint,aps,nofootinbib]{revtex4}
\usepackage[utf8]{inputenc}
\usepackage{graphicx}
\usepackage{latexsym,amsmath,amssymb,amsthm,lmodern,float,url}
\usepackage{natbib}
\usepackage{color}
\usepackage{microtype}
\usepackage{import}
\usepackage{bbold}
\usepackage[plain]{fancyref}
\usepackage{varioref}
\usepackage{slashed}
\usepackage{multirow}
\usepackage{tikz}
\usepackage{scrextend}
\usepackage{braket}
\usetikzlibrary{shapes}
\usetikzlibrary{positioning}
\usepackage[normalem]{ulem}
\usepackage{multirow}
\usepackage{makecell}
\usepackage{comment}
\usepackage{todonotes}
\usepackage{tabularray}
\usepackage{float}
\allowdisplaybreaks

\newcommand{\eqs}[1]{\begin{align} #1 \end{align}}

\newcommand{\fig}[1]{Fig.~\ref{fig:#1}}

\usepackage[colorlinks=true,backref=false, linktocpage=true,
citecolor=blue,urlcolor=blue,linkcolor=blue,pdfpagemode=UseOutlines]{hyperref}
\hypersetup{%
  bookmarksnumbered=true,
  pdftitle = {},
  pdfsubject = {},
  pdfauthor = {},
  pdfkeywords = {}
}

\let\Im\undefined

\DeclareMathOperator{\Im}{Im}
\def\be{\begin{equation}}
\def\ee{\end{equation}}
\def\bea{\begin{eqnarray}}
\def\eea{\end{eqnarray}}

\newcommand{\nocontentsline}[3]{}
\newcommand{\tocless}[2]{\bgroup\let\addcontentsline=\nocontentsline#1{#2}\egroup}

\DeclareRobustCommand{\orcidicon}{\hspace{-2.1mm}
\begin{tikzpicture}
\draw[lime,fill=lime] (0,0.0) circle [radius=0.13] node[white] {{\fontfamily{qag}\selectfont \tiny ID}}; \draw[white,fill=white] (-0.0525,0.095) circle [radius=0.007]; 
\end{tikzpicture} \hspace{-3.7mm} }
\foreach \x in {GC} {\expandafter\xdef\csname orcid\x\endcsname{\noexpand\href{https://orcid.org/0000-0001-5761-9590} {\noexpand\orcidicon}}}

\begin{document}

\title{ Entanglement Maximization and Mirror Symmetry in \\ 
Two-Higgs-Doublet Models}

\author{\vspace{0.2cm}
{Marcela Carena$^{\,1,2,3,4}$, Guglielmo Coloretti$^5$, Wanqiang Liu$^{\,2,3,4}$,\\  Mira Littmann$^2$, Ian Low$^{\,6,7}$, Carlos E.~M.~Wagner$^{\,1,2,3,6}$}
 }
\affiliation{\vspace*{0.2cm}
$^1$\mbox{\small Perimeter Institute for Theoretical Physics, 31 Caroline St. N., Waterloo, Ontario N2L 2Y5, Canada}\\
$^2$\mbox{\small Physics Depatrment and Enrico Fermi Institute, University of Chicago, Chicago, IL 60637, USA}\\
$^3$\mbox{\small Kavli Institute for Cosmological Physics, University of Chicago, Chicago, IL 60637, USA}\\
$^4$\mbox{\small Fermi National Accelerator Laboratory, Batavia,  Illinois, 60510, USA}\\
$^5$\mbox{\small Physik-Institut, Universit\"{a}t Z\"{u}rich, Winterthurerstrasse 190, CH–8057 Z\"{u}rich, Switzerland}\\
$^6$\mbox{\small High Energy Physics Division, Argonne National Laboratory, Argonne, IL 60439, USA}\\
$^7$\mbox{\small Department of Physics and Astronomy, Northwestern University, Evanston, IL 60208, USA} \\
}

\begin{abstract}
We consider  2-to-2 scatterings of Higgs bosons in a CP-conserving two-Higgs-doublet model (2HDM) and study the implication of maximizing the entanglement in the flavor space, where the  two  doublets $\Phi_a$,  $a=1,2$, can be viewed as a qubit: $\Phi_1=|0\rangle$ and $\Phi_2=|1\rangle$. More specifically, we compute the scattering amplitudes for $\Phi_a \Phi_b \to \Phi_c \Phi_d$ and require  the outgoing flavor entanglement to be maximal for a full product basis such as the computational basis, which consists of $\{|00\rangle,|01\rangle,|10\rangle,|11\rangle\}$. In the unbroken phase and turning off the gauge interactions, entanglement maximization results in the appearance of an $U(2)\times U(2)$ global symmetry among the quartic couplings, which in general is broken  softly by the mass terms. Interestingly,  once the Higgs bosons acquire vacuum expectation values, maximal entanglement enforces an exact  $U(2) \times U(2)$ symmetry, which is 
spontaneously broken to $U(1)\times U(1)$. As a byproduct, this gives rise to Higgs alignment as well as to the existence of 6 massless Nambu-Goldstone bosons. The  $U(2)\times U(2)$ symmetry can be gauged to lift the massless Goldstones, while maintaining maximal entanglement demands the presence of a discrete $\mathrm{Z}_2$ symmetry interchanging the two gauge sectors. The model is custodially invariant  in the scalar sector, and the inclusion of fermions requires a mirror dark sector, related to the standard one by the $\mathrm{Z}_2$ symmetry.
\end{abstract}




\maketitle

\tableofcontents
\newpage 

\section{Introduction}
\begin{flushright}
{\it Mirrors and parenthood are abominable,  since  they multiply  and extend \\ the visible Universe.}\\  Jorge Luis Borges, Tl\"on, Uqbar, Orbis Tertius (1940). \\
\end{flushright} 

Symmetry is one of the most powerful principles in physics. However, its existence in a physical system is either given as an input or emergent as a collective behavior. We do not have any organizing principle as to when and where a symmetry might appear. Given the unique and central role played by symmetry in nature, it is  surprising that very little has been known about its origin. 

Recently an intriguing correlation between symmetry and entanglement suppression was observed and studied in several systems, including non-relativistic baryon-baryon interactions \cite{Beane:2018oxh,Low:2021ufv,Liu:2022grf},  four-quark exotic mesons as shallow molecular bound states \cite{Hu:2024hex} and scattering of Higgs bosons in two-Higgs-doublet models (2HDMs) \cite{Carena:2023vjc}.  So far the symmetries involved are global internal or spacetime symmetries. Although we do not yet know whether symmetry can be the outgrowth of a more fundamental principle, these studies are suggestive and, at a minimum, open up the possibility of studying symmetry from the information-theoretic viewpoint. For a recent systematic investigation, see Ref.~\cite{McGinnis:2025brt}. Separately, entanglement minimization has also been invoked to deduce the flavor patterns in the Standard Model \cite{Thaler:2024anb}.

With the benefit of hindsight it is perhaps not surprising that there might be a relation between symmetry and entanglement suppression. Quantum entanglement is commonly quantified using entropic measures such as the von Neumann entropy or the R\'enyi entropy. In turn entanglement suppression is defined as vanishing entanglement power for a unitary operator \cite{Beane:2018oxh,Low:2021ufv} and is associated with  entropy suppression,  
which signals the presence of ``order'' -- smaller entropy implies less randomness and more order in the system. On the other hand, symmetry is in many ways also about the existence of order -- a system with more symmetries is also more orderly. The recent studies represent an important first step toward predictive power, in that one can make precise and quantitative statements on entanglement suppression and the resulting appearance of symmetry.

In this work we continue the exploration of entanglement and symmetry in 2HDMs initiated in Refs.~\cite{Carena:2023vjc,Kowalska:2024kbs,Chang:2024wrx}. In particular,  it was found in Ref.~\cite{Carena:2023vjc} that an entanglement suppressing $S$-matrix in the 2-to-2 scattering of a charged and a neutral Higgs bosons $\Phi^+_a\Phi^0_b\to \Phi^+_c\Phi^0_d$ gives rise to a maximal $SO(8)$ symmetry in the scalar potential,\footnote{More precisely, we require the $S$-matrix to be in the equivalent class of the Identity gate.}
\begin{align}
&V =m_{11}^2\Phi_1^\dagger\Phi_1+m_{22}^2\Phi_2^\dagger\Phi_2
-[m_{12}^2\Phi_1^\dagger\Phi_2+{\rm h.c.}]\nonumber+\frac 1 2\lambda_1(\Phi_1^\dagger\Phi_1)^2
+\frac 1 2\lambda_2(\Phi_2^\dagger\Phi_2)^2
\nonumber\\
&\qquad +\lambda_3(\Phi_1^\dagger\Phi_1)(\Phi_2^\dagger\Phi_2)
+\lambda_4(\Phi_1^\dagger\Phi_2)(\Phi_2^\dagger\Phi_1)
\nonumber\\
&\qquad +\left\{\frac 1 2\lambda_5(\Phi_1^\dagger\Phi_2)^2
+\big[\lambda_6(\Phi_1^\dagger\Phi_1)
+\lambda_7(\Phi_2^\dagger\Phi_2)\big]
\Phi_1^\dagger\Phi_2+{\rm h.c.}\right\}\,,
 \label{eq:pot}
\end{align}
where $\Phi_1$ and $\Phi_2$ are two doublet scalars $\Phi_i=(\Phi^+_i, \Phi^0_i)^T$, $i=1,2$, charged under electroweak $SU(2)_L\times U(1)_Y$ with hypercharge $Y=1$. For entanglement minimization it is important that we consider 
non-identical ``qubits.'' Otherwise the requirement of spin statistics for identical bosons, or the constraints from crossing symmetry, takes precedent and  renders entanglement minimization impossible. For instance, when scattering two identical neutrons in the low-energy when the $s$-wave dominates, Fermi-Dirac statistics  implies the $S$-matrix  always projects an unentangled incoming state into  the $^1S_0$ spin configuration, $(|\!\!\downarrow\uparrow\rangle - |\!\!\uparrow\downarrow\rangle)/\sqrt{2}$, which is anti-symmetric and maximally entangled. Thus the entanglement power of the $S$-matrix in the neutron-neutron scattering never vanishes. This is well-known in low-energy QCD and similar  observations were made on the 2HDM case in Ref.~\cite{Chang:2024wrx}.

The above considerations suggest it might be possible to achieve maximal entanglement in all scattering channels, without running foul with other prime principles. Moreover, just like minimal entanglement, maximal entanglement  also occupies a special corner in the parameter space of a theory, and one might wonder how that distinction manifests itself in the Lagrangian. We are also guided by the   intuition that a system with maximum randomness may look symmetric. For instance, a ferromagnet above the critical temperature where all the spins are randomly oriented in space is invariant under rotation. We will see that this is indeed borne out in 2HDMs, by demanding maximally entangled outgoing states in every scattering channel, for initial states that correspond to basis states in the computational basis: $\Phi_1\Phi_1, \Phi_1\Phi_2, \Phi_2\Phi_1,
\Phi_2\Phi_2$. The resulting symmetry is $U(2)\times U(2)$ before gauging.  In the unbroken phase, the symmetry is approximate in that it is active only among the quartic couplings and  broken softly by the masses. After the Higgs fields acquire vacuum expectation values (VEVs), maintaining maximal entanglement results in a $U(2)\times U(2)$ symmetry spontaneously broken to $U(1)\times U(1)$. In this case there are 6 massless Nambu-Goldstone bosons, similar to the situation in Ref.~\cite{Carena:2023vjc}. What is different here, is that it is possible to gauge the global symmetry in order to lift the mass of the Nambu-Goldstone bosons while still maintaining entanglement maximization. Moreover, a discrete $\mathrm{Z}_2$ symmetry emerges which enforces equal gauge coupling constants $g_{\sf{L}}=g_{\sf{R}}$. The Higgs field alignment, where the Higgs boson  couples to the $W$ and $Z$ bosons with full strengths \cite{Gunion:2002zf,Carena:2013ooa,Carena:2014nza,Carena:2015moc,Low:2020iua}, is a natural outcome in this model.

This work is organized as follows : In section \ref{sect:2} we describe the concept of flavor entanglement in 2HDM scattering states. In section \ref{sect:3} we discuss the question of  entanglement maximization in the unbroken phase, making emphasis on the enhanced $U(2) \times U(2)$ symmetry of the quartic interactions arising from this condition. In section \ref{sect:4} we generalize the study to the broken phase, in which the symmetry becomes exact, but spontaneously broken, leading to the presence of six massless Goldstone bosons. In section \ref{sect:5} we study the case of  gauging the  $SU(2)\times SU(2)$ to lift the mass of the Goldstone modes, showing that maximal entanglement is preserved in the presence of a discrete $\mathrm{Z}_2$ symmetry. which enforces the equality of both gauge couplings. We also point out the presence of custodial symmetry in the setup, which would protect the precision electroweak corrections at the tree level. Then in Section \ref{sect:loop} we discuss loop corrections and the inclusion of fermions in the 2HDM.  We reserve section \ref{sect:con} for our conclusions. In the Appendix \ref{sect:app} we provide explicit expressions for the scattering amplitudes in various scattering channels.

\section{Flavor Entanglement in 2-to-2 Scattering}
\label{sect:2}

Here we briefly summarize the key concepts of quantum entanglement  and their applications in 2-to-2 scattering.

We start with two distinguishable qubits, Alice ($A$) and Bob ($B$), each with its own basis vectors $\{|0\rangle_I, |1\rangle_I\}$, $I=A,B$.  For example,  we can choose $\{|0\rangle_A, |1\rangle_A\} = \{\Phi^+_1,  \Phi^+_2\}$ and $\{|0\rangle_B, |1\rangle_B\} = \{\Phi^0_1,  \Phi^0_2\}$, where the distinguishability comes from the different electroweak quantum numbers in the qubits. It is common to define the computational basis  $\{|00\rangle,|01\rangle,|10\rangle,|11\rangle\}$, where $|ij\rangle = |i\rangle_A\otimes |j\rangle_B$. A  state vector in the computational basis can then be written as $|\psi\rangle= \alpha |00\rangle +\beta |01\rangle +\gamma|10\rangle + \delta |11\rangle$, with the normalization $|\alpha|^2+|\beta|^2+|\gamma|^2+|\delta|^2=1$.

There are several quantitative measures of entanglement in two-qubit systems  \cite{Kraus_2001}. Some common examples are the von Neumann entropy $E_{vN}({\rho_A}) = -\mathrm{Tr}({\rho}_A\ln{\rho}_A)$ and the linear entropy $E_L({\rho_A}) = 1-\mathrm{Tr}({\rho}_A^2)$, where ${\rho}=|\psi\rangle\langle\psi|$ is the density matrix and ${\rho}_{A/B}={\rm Tr}_{B/A}( {\rho})$ is the reduced density matrix for Alice/Bob. Ref.~\cite{Low:2021ufv} showed that, for two-qubit pure states, all entanglement measures  are related to the concurrence $\Delta$ \cite{PhysRevLett.78.5022,PhysRevLett.80.2245}, which is defined as  
\be
\label{eq:con}0\quad \le  \quad 
\Delta(|\psi\rangle) \equiv 2|\alpha\delta-\beta\gamma|\le \quad 1 \ ,
\ee
such that the linear entropy $E_L(\rho_{A/B})= \Delta^2/2 $. The concurrence has a minimum at 0
such that $E_L(\rho_{A/B})=0 $ if and only if $|\psi\rangle$ is a product state and unentangled, $|\psi\rangle = |\psi_A\rangle\otimes |\psi_B\rangle$.  If $|\psi\rangle $ is a maximally entangled state such as the  Bell states:
\be\label{eq:Bell}
\ket{\Phi^{\pm}}=\frac1{\sqrt{2}}(|00\rangle \pm |11\rangle) \ , \qquad  \ket{\Psi^{\pm}}=\frac1{\sqrt{2}}(|10\rangle \pm |01\rangle) \ ,
\ee
$\Delta(|\psi\rangle)$ reaches the maximum at 1.

In this work we would like to  investigate the information-theoretic property of 2-to-2 scatterings in 2HDMs, focusing on  entanglement in the flavor space $\Phi_a\Phi^\prime_b\to \Phi_c\Phi^\prime_d$. In terms of Alice and Bob qubits, we identify $\{|0\rangle_A,|1\rangle_A\}=\{\Phi_1, \Phi_2\}$ and $\{|0\rangle_B,|1\rangle_B\}=\{\Phi^\prime_1, \Phi^\prime_2\}$. The $S$-matrix of the scattering process  is related to the transition $T$-matrix, 
\be
\label{eq:Smat}
S= 1 + i \,T\ ,
\ee
whose  matrix element is
\be
\label{eq:ampT}
 \langle \Phi_c\Phi^\prime_d |\, iT\, |\Phi_a\Phi^\prime_b\rangle  = i (2\pi)^4 \delta^{(4)}(p_a+p_b-p_c-p_c)  M_{ab,cd} \ .
\ee
$\cal M_{ab,cd}$ is the scattering amplitude one typically computes in perturbation theory. In the previous work we focused on the entanglement power of the $S$-matrix \cite{Carena:2023vjc}, which is a unitary operator and can be interpreted as a two-qubit quantum logic gate. In the current study we will instead focus on the entanglement property of the scattering amplitude, following a similar spirit as in Refs.~\cite{Cervera-Lierta:2017tdt,Miller:2023ujx,Bai:2023tey}. The difference being that the $T$-matrix, and therefore the amplitude itself, is not a unitary operator. In fact,  the optical theorem states
\be
\label{eq:Tmatrix}
i(T^\dagger - T) =TT^\dagger\ ,
\ee
which follows from the unitarity of the $S$-matrix. At the tree-level, the amplitude does not have an imaginary part and the amplitude is Hermitian. 

It is worth commenting on why we choose to study the amplitude, rather than the $S$-matrix itself, when it comes to entanglement maximization. The Identity operator in Eq.~(\ref{eq:Smat}) gives the "nothing happens" part of the collision where the two particles pass through each other without interacting. The Identity also suppresses entanglement because it always takes a product state to another product state and never generates entanglement. In a weakly coupled theory, the $T$-matrix is suppressed by the small coupling constant so the Identity is always the dominant part of the $S$-matrix. Therefore, we expect the $S$-matrix will never maximize entanglement for as long as the perturbative expansion remains valid.\footnote{This argument also makes it clear that, in order for the $S$-matrix to be a SWAP gate, which also suppresses entanglement, the theory must be strongly interacting \cite{Low:2021ufv}.}

The $T$-matrix, although not a unitary operator,  gives the probability of the initial state scattering into a particular flavor and kinematic configuration. Therefore the entanglement property of the scattered states often depends on the particular location in the kinematic phase space. Indeed, in any given theory one should not be surprised to find maximally entangled states at special $(\theta,\phi)$ angles in the phase space \cite{Cervera-Lierta:2017tdt,Miller:2023ujx,Bai:2023tey}; these are generic features of scattering events. What is less generic, which might hint at something special such as a symmetry, is that the outgoing states are maximally entangled states in all corners of phase space, irrespective of the kinematics. 

To put it differently, the scattered state $T\, |\Phi_a\Phi^\prime_b\rangle$ in general is a linear superposition of all possible outcomes and, therefore, an entangled state between the flavor quantum number and the kinematic variables. The entanglement maximization we will demand amounts to requiring that the outgoing state is a product state between flavor and momentum, and that the wave function in the flavor subspace is maximally entangled:
\be 
\label{eq:disent}
T\, |\Phi_a\Phi^\prime_b\rangle = |\mbox{flavor}\rangle \otimes |\mbox{kinematics}\rangle \ , \qquad \Delta(|\mbox{flavor} \rangle)  =1.
\ee

With this special property,  the reduced density matrix in the flavor space remains a pure state after tracing out the kinematic variables. 
The same consideration applies to previous works on entanglement suppression and emergent symmetries -- when the $S$-matrix suppresses entanglement and exhibits enhanced symmetries, the reduced density matrix after tracing out the kinematics is that of a pure state.

\section{ Entanglement Maximization in the Unbroken Phase}
\label{sect:3}

In this section, we discuss the conditions under which the 2-to-2 scattering, $\Phi_a\Phi^\prime_b\to \Phi_c\Phi^\prime_d$,  produces maximal entanglement in flavor space, when the $U(2)$ symmetry is not spontaneously broken. For simplicity we assume CP conservation and the $\lambda_i$'s in Eq.~(\ref{eq:pot}) to be real parameters, although our results can be easily generalized to the CP-violating case. We turn off gauge interactions in this section and will re-introduce them when discussing the broken phase.  We define the flavor structure of the amplitudes as four by four matrices in the basis of states corresponding to the computational basis,
$\Phi_1\Phi_1,\Phi_1\Phi_2,\Phi_2\Phi_1,\Phi_2\Phi_2$.  
The flavor structure of the scattered states is given by
\begin{equation}
    |\Phi_c\Phi_d \rangle =  i {\cal M}_{ab,cd} \  |\Phi_a\Phi_b \rangle
\end{equation}
with ${\cal M}_{ab,cd}$ being the scattering amplitudes in flavor space.

In the unbroken phase, the scattering is induced by contact quartic couplings in Eq.~(\ref{eq:pot}). The scattering amplitudes in various channels are given by \cite{Kowalska:2024kbs}:
\bea \label{eq::polish}
  i{\cal M}_0^{(1)}(\Phi^+\Phi^0\to \Phi^+\Phi^0) &=& 
-i \begin{bmatrix}   
 \lambda_1 & \lambda_6 & \lambda_6 & \lambda_5 \\[-.2cm]
     \lambda_6 & \lambda_{3} & \lambda_{4} &  \lambda_7 \\[-.2cm]
       \lambda_6 & \lambda_{3} & \lambda_{4} &  \lambda_7 \\[-.2cm]
     \lambda_5 & \lambda_7 &  \lambda_7 &  \lambda_2
     \end{bmatrix} \ , \\
       i{\cal M}_0^{(2)}(\Phi^+\tilde{\Phi}^0\to \Phi^+\tilde{\Phi}^0) &=& 
-i \begin{bmatrix}   
 \lambda_1 & \lambda_6 & \lambda_6 & \lambda_4 \\[-.2cm]
     \lambda_6 & \lambda_{3} & \lambda_{5} &  \lambda_7 \\[-.2cm]
       \lambda_6 & \lambda_{5} & \lambda_{3} &  \lambda_7 \\[-.2cm]
     \lambda_4 & \lambda_7 &  \lambda_7 &  \lambda_2
     \end{bmatrix} \ , \\
       i{\cal M}_0^{(3)}(\Phi^+\Phi^-\to \Phi^+\Phi^-) &=& i{\cal M}_0^{(3)}(\Phi^0\tilde{\Phi}^0\to \Phi^0\tilde{\Phi}^0) = 
     -i \begin{bmatrix}   
 2\lambda_1 & 2\lambda_6 & 2\lambda_6 & \lambda_{34} \\[-.1cm]
    2 \lambda_6  & \lambda_{34} & 2\lambda_{5} & 2 \lambda_7 \\[-.1cm]
      2 \lambda_6 & 2 \lambda_{5} &\lambda_{34} & 2 \lambda_7 \\[-.1cm]
     \lambda_{34} & 2\lambda_7 &  2\lambda_7 &  2\lambda_2
     \end{bmatrix} \ , \\
      i{\cal M}_0^{(4)}(\Phi^0\tilde \Phi^0\to \Phi^+\Phi^-) &=&  i{\cal M}_0^{(4)}(\Phi^+\Phi^-\to \tilde{\Phi}^0\Phi^0) 
 = -i \begin{bmatrix}   
 \lambda_1 & \lambda_6 & \lambda_6 & \lambda_3 \\[-.2cm]
     \lambda_6 & \lambda_{4} & \lambda_{5} &  \lambda_7 \\[-.2cm]
       \lambda_6 & \lambda_{5} & \lambda_{4} &  \lambda_7 \\[-.2cm]
     \lambda_3 & \lambda_7 &  \lambda_7 &  \lambda_2
     \end{bmatrix} \ ,
     \label{eq::polish1}
\eea
where $\lambda_{34}=\lambda_3+\lambda_4$ and $\tilde{\Phi}^0=(\Phi^0)^*$. We see there are five channels with differing charge assignments  of the scattering states. The matrices are written in the flavor basis $\{|00\rangle, |01\rangle, |10\rangle,|11\rangle\}$. 

We start with initial states whose flavor quantum numbers are the basis states in the computational basis, $\{|00\rangle, |01\rangle, |10\rangle,|11\rangle\}$, and study the conditions under which the flavor entanglement in the outgoing state is maximized for every basis state and for every channel $i = 1,..,4$,
\be
\Delta(|j\rangle) = \Delta({\cal M}^{(I)}_0 |i\rangle)=1\ , \qquad \forall\   |i\rangle = \{|00\rangle, |01\rangle, |10\rangle,|11\rangle\} \ ,
\ee
where the concurrence $\Delta$ is computed using Eq.~(\ref{eq:con}). For instance, let's consider the charged-neutral channel in $\Phi^+\Phi^0\to \Phi^+\Phi^0$:
\bea
\Delta({\cal M}_0^{(1)}|00\rangle) &=& \frac{ \left| 2 ( \lambda_1 \lambda_5 - \lambda_6^2)\right|}{2 \lambda_6^2 + \lambda_1^2 + \lambda_5^2}\ ,\\
\Delta({\cal M}_0^{(1)}|01\rangle)&=&\Delta({\cal M}_0^{(1)}|10\rangle)=
\frac{ \left| 2 ( \lambda_6 \lambda_7 - \lambda_3 \lambda_4)\right|}{\lambda_7^2+ \lambda_6^2 + \lambda_3^2 + \lambda_4^2}\ ,\\
\Delta({\cal M}_0^{(1)}|11\rangle)&=&\frac{ \left| 2 ( \lambda_2 \lambda_5 - \lambda_7^2)\right|}{2 \lambda_7^2 + \lambda_2^2 + \lambda_5^2}\ .
\eea
Demanding the concurrence is maximized, $\Delta=1$, leads to the following two families of solutions 
\begin{eqnarray}
   &{\rm (i)}&\quad \lambda_1 = \pm \lambda_5\ , \quad 
    \lambda_3 = \pm \lambda_4 \ ,\quad
    \lambda_2 = \pm \lambda_5\ ,\quad \lambda_6 = \lambda_7 = 0 \ ; 
    \label{lambdas}\\
    &{\rm (ii)}& \quad\lambda_1=\lambda_2=\lambda_3=\lambda_4=\lambda_5=0 \ ,\quad \lambda_6=\pm \lambda_7 \ .
\end{eqnarray}
Further demanding entanglement maximization for all charged/neutral channels, ${\cal M}^{(I)}$, $i=1,\cdots,4$, we obtain three families of solutions:
\bea
\label{eq:sol1}
\mbox{(1)}&:& \lambda_1=\lambda_2=\lambda_3=\lambda_4=\pm \lambda_5\neq 0 \ , \quad \lambda_6=\lambda_7=0\ ,\\
\mbox{(2)}&:& \lambda_1=\lambda_2=-\lambda_3=-\lambda_4=\pm \lambda_5\neq 0 \ , \quad \lambda_6=\lambda_7=0\ ,\\
\mbox{(3)}&:& \lambda_1=\lambda_2=\lambda_3=\lambda_4=\lambda_5= 0 \ , \quad \lambda_6=\pm \lambda_7\neq 0\ ,
\eea
Among these  solutions, Solution (1) with $\lambda_4=\lambda_5$ and Solution (3) with $\lambda_6=\lambda_7$ agree with those two presented in Table 2 of Ref.~\cite{Kowalska:2024kbs}. The other solutions are new. However, it turns out that Solutions (3) and Solution (2) give rise to a scalar potential that is not bounded from below. Indeed, the necessary conditions to avoid a bounded from below potential for
$\lambda_{1,2} \geq 0$ and real couplings are \cite{Bahl:2022lio}:
\begin{eqnarray}
\sqrt{\lambda_1 \lambda_2 } + \lambda_3 + \lambda_4 + \lambda_5 - | \lambda_6 + \lambda_7| & \geq & 0
\ ,\\
\sqrt{\lambda_1 \lambda_2 } + \lambda_3 + \lambda_4 - \lambda_5 
& \geq & 0\ .
\end{eqnarray}
Henceforth we focus on Solutions (1).

It is informative to look at the scattering amplitudes for Solution (1):
\bea
\label{eq:mm1}
{\cal M}_0^{(1)}&\sim& \begin{bmatrix} 
 1 & 0 & 0 & \pm 1 \\[-.25cm]
     0 & 1 &  1 &  0 \\[-.25cm]
      0 &  1 & 1 &  0 \\[-.25cm]
   \pm 1 & 0 &  0 & 1
     \end{bmatrix} \ , \\
     \label{eq:mm2}
{\cal M}_0^{(2)}&\sim& {\cal M}_0^{(3)}\sim {\cal M}_0^{(4)}\sim
\begin{bmatrix}   
 1 & 0 & 0 & 1 \\[-.25cm]
     0 & 1 & \pm 1 &  0 \\[-.25cm]
      0 & \pm 1 & 1 &  0 \\[-.25cm]
    1 & 0 &  0 & 1
     \end{bmatrix}\ ,
\eea
where the $\pm$ signs correspond to $\lambda_i=\pm\lambda_5$, $i=1,\cdots, 4$.
From the above we clearly see that the scattering produces a maximally entangled Bell state in Eq.~(\ref{eq:Bell}) for every incoming basis states in the computational basis. For generic initial product states, we discuss the entanglement properties in Appendix \ref{app:gen_ent}.

The quartic terms in the scalar potential that achieve the maximal entanglement solution in Eq.~(\ref{eq:sol1}) can now be written as  
\begin{align}
\label{eq:vp4}
\lambda=\lambda_i=\lambda_5&: \quad  V^{(4)}_+ = \frac \lambda 2 \left[(\Phi^\dagger\Phi)^2+(\Phi^\dagger\tau_1\Phi)^2\right]\ , \\
\label{eq:vm4}
\lambda=\lambda_i=-\lambda_5&:  \quad  V^{(4)}_- = \frac \lambda 2 \left[(\Phi^\dagger\Phi)^2+(\Phi^\dagger\tau_2\Phi)^2\right]\ ,
\end{align}
where $\Phi=(\Phi_1,\Phi_2)^T$ and $\tau^a$, $a=1,2,3$, are the Pauli matrices acting on the flavor space. To identify the symmetries of $V_\pm$, we observe that $(\Phi^\dagger\Phi)^2$ is invariant under an $U(4)$ symmetry under which $\Phi$ is in the fundamental representation. The generators of the $U(4)$ symmetry are $\{\openone\otimes \openone, \openone\otimes \sigma^b, \tau^a\otimes \openone, \tau^a\otimes\sigma^b\}$, $a,b=1,2,3$. Here $\tau^a$ and $\sigma^b$ are the Pauli matrices acting on the flavor space and the weak isopsin space, respectively. In addition, $\openone$ is a $2\times 2$ identity matrix. The second term in $V_\pm$, however, only preserves a subgroup of $U(4)$. In fact, one can read off the generators of $U(4)$ which are preserved by $V_\pm$:
\bea
\label{eq:vplus}
V^{(4)}_+&:& \quad \{ \openone\otimes \sigma^\mu,  \tau^1\otimes\sigma^\mu\} \ , \\
\label{eq:vminus}
V^{(4)}_-&:& \quad \{\openone\otimes \sigma^\mu,  \tau^2\otimes\sigma^\mu\} \ ,
\eea
where $\mu=0,\cdots, 3$ and $\sigma^0=\openone$. That is, in each case there are eight generators whose actions  leave the potential invariant. These generators factorize into two (different) copies of $U(2)$'s:
\bea 
\label{eq:sym1}
V^{(4)}_+&:& U(2)_S=T_S^\mu=\frac{\openone+\tau^1}{2}\otimes \sigma^\mu \ , \quad U(2)_A=T_A^\mu=\frac{\openone-\tau^1}{2}\otimes \sigma^\mu \ ,\\
\label{eq:sym2}
V^{(4)}_-&:& U(2)_L=T_L^{\mu}=\frac{\openone+\tau^2}{2}\otimes \sigma^\mu \ , \quad U(2)_R=T_R^{\mu}=\frac{\openone-\tau^2}{2}\otimes \sigma^\mu \ .
\eea
 It is easy to see that the product $T_S^\mu T_A^\nu=T_L^\mu T_R^\nu=0$ and that the electroweak $SU(2)_L$ is the direct sum: $SU(2)_{S+A}=SU(2)_{R+L}$. We conclude that, after imposing entanglement maximization conditions, the quartic couplings in the scalar potential exhibit enhanced symmetries $U(2)_S\times U(2)_A$ or $U(2)_L\times U(2)_R$.

The symmetries in Eqs.~(\ref{eq:sym1}) and (\ref{eq:sym2}) suggest the following bases in the flavor space,
\begin{align}
    \Phi_\mathsf{S,A}&= \frac{1}{\sqrt{2}}(\Phi_1\pm \Phi_2), \label{eq:phias}
    \\
    \Phi_\mathsf{L,R}&= \frac{1}{\sqrt 2}(\Phi_1\mp i\Phi_2)\label{eq:philr} ,
\end{align}
since $(\openone\pm \tau^1)/2$ are the projection operators from $(\Phi_1,\Phi_2)$ basis to $(\Phi_\mathsf{S},\Phi_\mathsf{A})$ basis, and similarly $(\openone\pm \tau^2)/2$ for the $(\Phi_\mathsf{L},\Phi_\mathsf{R})$ basis. In the new bases the scalar potential become
\begin{align}
    V^{(4)}_+&=   \lambda  \left[(\Phi_{\sf{S}}^\dagger\Phi_{\sf{S}})^2+(\Phi_{\sf{A}}^\dagger\Phi_{\sf{A}})^2\right]\ ,\\
    V^{(4)}_-&=   \lambda  \left[(\Phi_{\sf{L}}^\dagger\Phi_{\sf{L}})^2+(\Phi_{\sf{R}}^\dagger\Phi_{\sf{R}})^2\right]\ .
\end{align}
The two sectors $\{\Phi_{\sf{S/L}},\Phi_{\sf{A/R}}\}$ decouple from each other in $V_\pm$. As such, the full quartic amplitude can be written as the sum of the subamplitudes from the two decoupled sectors.

The above observations allow us to understand the particular patterns exhibited in Eqs.~(\ref{eq:mm1}) and (\ref{eq:mm2}). When we scatter two flavor qubits $\mathsf{a}$ and $\mathsf{b}$ in the $\lambda_i=\lambda_5$ solution, the quartic amplitude is the sum of the two subamplitudes: one in the $\Phi_{\sf{S}}\Phi_{\sf{S}}$ scattering and the other  in the $\Phi_{\sf{A}}\Phi_{\sf{A}}$ scattering. Therefore,  the total amplitude can be factorized as 
\bea\label{eq:XX1}
\mathcal{M}^{(I)}_0&\propto& \frac{\openone+\tau^1}{2} \otimes \frac{\openone+\tau^1}{2}+\frac{\openone-\tau^1}{2} \otimes \frac{\openone-\tau^1}{2}\nonumber\\
&=&\frac{\openone\otimes\openone+\tau^1\otimes\tau^1}{2} = \frac12\, \begin{bmatrix} 
 1 & 0 & 0 &  1 \\[-.3cm]
     0 & 1 &  1 &  0 \\[-.3cm]
      0 &  1 & 1 &  0 \\[-.3cm]
   1 & 0 &  0 & 1
     \end{bmatrix} \ .
\eea
The $T$-matrix projects the flavor amplitude to the eigenbasis of $\tau^1\otimes\tau^1$ with the eigenvalue $+1$. This property is reminiscent of the repetition code in quantum error correction. We discuss this connection in Appendix \ref{app:repetition_code}.

 In the $\lambda_i=-\lambda_5$ solution,  due to the appearance of the additional $i$ in the definition of $\Phi_{\sf{L/R}}$ in Eq.~(\ref{eq:philr}), the $\sf{L/R}$ assignment of the CP-even particles are consistent with the fields, while the CP-odd ones are the opposite. Thus, for the scattering among the CP-even particles, i.e. the charge sectors (1) and (2), the  amplitudes ${\cal M}_0^{(1)}$ are written as as
\bea
\label{eq:YY1+}
\mathcal{M}^{(1)}_0&\propto& \frac{\openone+\tau^2}{2} \otimes \frac{\openone+\tau^2}{2}+\frac{\openone-\tau^2}{2} \otimes \frac{\openone-\tau^2}{2}\nonumber\\
&=&\frac{\openone\otimes\openone+\tau^2\otimes\tau^2}{2} = \frac12\begin{bmatrix} 
 1 & 0 & 0 &  -1 \\[-.3cm]
     0 & 1 &  1 &  0 \\[-.3cm]
      0 &  1 & 1 &  0 \\[-.3cm]
   -1 & 0 &  0 & 1
     \end{bmatrix} \ ,
\eea
while for the scatterings between CP-even and CP-odd particles, i.e. the charge sectors (2-4), the quartic amplitudes are
\bea
\label{eq:YY1-}
\mathcal{M}^{(2/3/4)}_0&\propto& \frac{\openone+\tau^2}{2} \otimes \frac{\openone-\tau^2}{2}+\frac{\openone-\tau^2}{2} \otimes \frac{\openone+\tau^2}{2}\nonumber\\
&=&\frac{\openone\otimes\openone-\tau^2\otimes\tau^2}{2} = \frac12\begin{bmatrix} 
 1 & 0 & 0 &  1 \\[-.3cm]
     0 & 1 &  -1 &  0 \\[-.3cm]
      0 &  -1 & 1 &  0 \\[-.3cm]
   1 & 0 &  0 & 1
     \end{bmatrix} \ .
\eea


\section{Entanglement Maximization  in the Broken Phase}
\label{sect:4}

In the broken phase of 2HDM where $SU(2)_L\times U(1)_Y$ is broken down to $U(1)_{em}$,  there are  two scalar vacuum expectation values (VEVs) in general, $v_1$ and $v_2$, which can be made real and nonnegative and $(v_1^2+v_2^2)^{1/2}\equiv v= 246$ GeV.  We also define $t_\beta = {v_2}/{v_1} \ge 0$, $0\le \beta \le \pi/2$, such that $c_\beta\equiv \cos\beta=v_1/v$ and $s_\beta\equiv\sin\beta=v_2/v$. To discuss the phenomenological property of the lightest CP-even Higgs boson upon spontaneous symmetry breaking, it will be helpful to perform a global $U(2)$ rotation of $\Phi=(\Phi_1, \Phi_2)^T$,  $\Phi \to {\cal H}\equiv (H_1,H_2)^T= {\cal U}\, \vec{\Phi}$, where ${\cal U}\in U(2)$, such that only one of the doublets receives a VEV: $\langle H_1^0 \rangle = {v}/{\sqrt{2}}$ and $\langle H_2^0 \rangle = 0$. This is the so-called Higgs basis \cite{Botella:1994cs} and a convenient choice for considering the alignment limit of the 125 GeV Higgs boson \cite{Carena:2013ooa,Carena:2014nza,Carena:2015moc,Low:2020iua}. The global $U(2)$ rotation leaves the scalar kinetic term invariant and can be viewed as a field redefinition which keeps the physics intact. Although the parameters appearing in Eq.~(\ref{eq:pot}) are not invariant under  $U(2)$ rotations, two potentials whose parameters are related by a $U(2)$ rotation are physically equivalent.  In the Higgs basis the scalar potential has the same form as in Eq.~(\ref{eq:pot}) but with the coefficients $\{m_1^2, m_2^2, m_{12}^2\}\to \{Y_1, Y_2, Y_3\}$ and $\lambda_i\to Z_i$.  Upon gauging $SU(2)_L\times U(1)_Y$, the 125 GeV Higgs boson is ``Standard Model-like'' when the Higgs basis coincides with the mass eigenbasis, which is ensured by the condition $Z_6=0$ \cite{Carena:2013ooa,Carena:2014nza,Carena:2015moc,Low:2020iua}.

\begin{figure}
    \centering
    \includegraphics[width=1\linewidth]{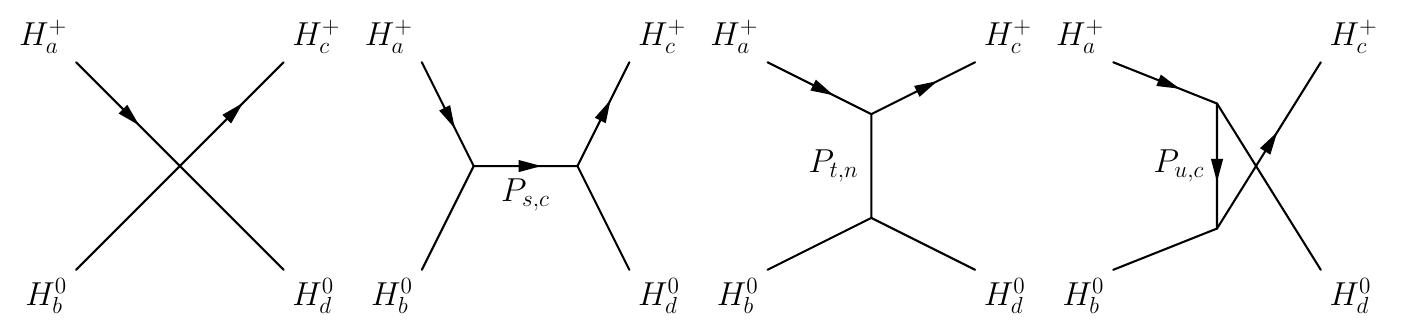}
    \caption{Feynman diagrams for $H^0 H^+ \to H^0 H^+$ scattering in the broken phase. $P_{r,k}$ corresponds to the propagators, where channels are labeled with $r \in \{s,t,u\}$, elements of the computational basis with $\{a,b,c,d\}\in\{0,1\}$ and masses with $c \in \{G^\pm, H^\pm \}$ and $n \in \{h,G^0,H,A\}$.}
    \label{fig:cnconc}
\end{figure}

In the broken phase, the 2-to-2 scattering of Higgs bosons at the tree-level contains four contributions:  the quartic and the $s/t/u$-channels, as exemplified in Fig.~\ref{fig:cnconc} for the charged-neutral sector. Notice that the kinematic dependence of each diagram in Fig.~\ref{fig:cnconc} is different and entanglement maximization requires the flavor and the kinematic quantum numbers of the outgoing state to disentangle as in Eq.~(\ref{eq:disent}). This implies it is not sufficient to demand that each contribution in Fig.~\ref{fig:cnconc} gives rise to a maximally entangled  state, but that they must result in {\em the same} maximally entangled state. Otherwise when summing over all four diagrams the final outgoing state will not be maximally entangled. Among the four diagrams, the contribution from the quartic coupling is identical with that computed in the unbroken phase. So our starting point is Solution (1) in Eq.~(\ref{eq:sol1}), $Z=Z_i=\pm Z_5$, which  miraculously results in the other three diagrams giving the same maximally entangled Bell state as in the quartic diagram.

Given that Solution (1) requires $Z_6=0$, we are automatically in the alignment limit and the Higgs basis is the mass eigenbasis, which simplifies the calculation significantly. More specifically, in the Higgs basis we have $H_1 =  (G^+, v/\sqrt{2}+H^0_1)^T$ and $H_2 = ( H^+, H^0_2)$, where $G^+$ is the charged Goldstone boson and $H^+$ is the charged scalar; they are both mass eigenstates. There are four additional mass eigenstates in the neutral sector: $(h, H, G^0, A)$: $h= \sqrt{2}~{\rm Re}[H_1^0]$ is the lightest CP-even scalar, which we assume to be the 125 GeV Higgs boson, $H=\sqrt{2}~{\rm Re}[H_2^0]$ and $A=\sqrt{2}~{\rm Im}[H_2^0]$ are the CP-even and CP-odd neutral scalars, respectively, and $G^0=\sqrt{2}~{\rm Im}[H_1^0]$ is a neutral Goldstone boson. In the Higgs basis, the minimization condition leads to the quadratic coefficients $Y_1=-Z_1v^2/2=-Zv^2/2$ and $Y_3=-Z_6 v^2/2=0$, while the mass matrices of the charged and CP even/odd neutral scalars are given by \cite{Carena:2023vjc}
\eqs{\label{eq:masq}
	M^2_{\pm} &= \begin{bmatrix} 0 & 0 \\ 0 & Y_2+Z_3v^2/2 \end{bmatrix} = \begin{bmatrix} 0 & 0 \\ 0 & Y_2+Z v^2/2 \end{bmatrix},\\ 
    \label{eq:mcpe}
	M^2_{\rm CP \ even} &= \begin{bmatrix} Z_1 v^2 & Z_6 v^2 \\ Z_6 v^2 & Y_2+(Z_3+Z_4+Z_5)v^2/2 \end{bmatrix}=\begin{bmatrix} Z v^2 & 0 \\0 & Y_2+(2Z+Z_5)v^2/2 \end{bmatrix} ,\\
        \label{eq:mcpo}
	M^2_{\rm CP \ odd} &= \begin{bmatrix} 0 & 0 \\ 0 & Y_2+(Z_3+Z_4-Z_5)v^2/2 \end{bmatrix}=\begin{bmatrix} 0 & 0 \\ 0 & Y_2+(2Z-Z_5)v^2/2 \end{bmatrix} \ ,
}
where we have plugged in $Z=Z_i$, $i=1,\cdots,4$, in the second equality.
In addition, the Feynman rules can be found in, for example, Ref.~\cite{Carena:2023vjc}. The full amplitude in each of the scattering channel $I$ is \eqs{\label{eq:amp_broken}
i{\cal M}^{(I)} &=  i{\cal M}_0^{(I)} - \frac{v^2}{2}\sum_{k}\sum_{r=s,t,u} {\cal M}^{(I)}_{r,k}\ P_{r,k} \ , 
 } 
where the propagators entering the $s/t/u$-channel diagrams are $P_{r,k} = i/(r-m_k^2)$, $r=s,t,u$.  Here masses in the propagator run through $m_k=\{m_h, m_H, m_{G^+}=0, m_A,m_{H^\pm}, m_{G^0}=0\}$, depending on the scattering channel. The quartic amplitudes ${\cal M}_0^{(I)}$ are given in Eqs.~(\ref{eq::polish})-(\ref{eq::polish1}), with the replacement $\lambda_i\to Z_i$. The trilinear vertices $M_{abc}$ and $M_{abc,0}$ are given by:
\begin{equation}
    \frac{\partial V}{\partial v}\Bigg\rvert_{v=0}=\sum_{a,b,c}\left[ M_{abc}H^+_a H^0_b H^-_c + M_{abc,0} H^0_a H^0_b H^{0,*}_c + \rm{h.c.} \right] \ .
\end{equation}
%

As an illustration we show the amplitude ${\cal M}^{(1)}$ for $H^+ H^0 \to H^+ H^0$. The trilinear vertices yield the following matrix elements (see Fig.~(\ref{fig:cnconc})):
\begin{align}
(\mathcal{M}^{(1)}_{s,k})_{ab,cd}&=M_{abk}M^*_{cdk} \ , \nonumber\\ (\mathcal{M}^{(1)}_{u,k})_{ab,cd} &= M_{adk}M^*_{cbk}\ , \\ (\mathcal{M}^{(1)}_{t,k})_{ab,cd} &= \sum_{i,j}\mathcal{R}_{ki}M_{aic}(\mathcal{R}_{kj}M_{djb,0})^*\nonumber \ .
\end{align}
with\footnote{We adopt a CP-conserving potential ($\Im[Z_5]=0$) and the symmetries considered in this article imply $Z_6=0$. Therefore, there is no mixing among the neutral fields.}
\begin{equation}
    \begin{bmatrix}
        h \\ H \\ G^0 \\ A
    \end{bmatrix} = \sqrt{2}~\mathcal{R} \begin{bmatrix}
        H^0_1 \\ H^0_1{}^* \\ H^0_2 \\ H^0_2{}^*
    \end{bmatrix}, \
     \mathcal{R} = \frac{1}{2}\begin{bmatrix}
        1 & 1 & 0 & 0 \\
        0 & 0 & 1 & 1 \\
        -i & i & 0 & 0 \\
        0 & 0 & -i & i
    \end{bmatrix} \ .
\end{equation}
Setting $Z=Z_i=\pm Z_5$ as in Eq.~(\ref{eq:sol1}), we obtain:
\begin{align}
\label{eq:mmmm1}
   {\cal M}^{(1)}_0 &= Z
   \begin{bmatrix}
    1 & 0 & 0 & \pm 1 \\[-.3cm]
    0 & 1 & 1 & 0 \\[-.3cm]
    0 & 1 & 1 & 0 \\[-.3cm]
    \pm 1 & 0 & 0 & 1 
   \end{bmatrix} \ ; & & \\
   \label{eq:m2s}
   {\cal M}^{(1)}_{s, G^\pm}&= Z^2
   \begin{bmatrix}
    1 & 0 & 0 & \pm 1 \\[-.3cm]
    0 & 0 & 0 & 0 \\[-.3cm]
    0 & 0 & 0 & 0 \\[-.3cm]
    \pm 1& 0 & 0 & 1 \\
   \end{bmatrix}\ ,
   &
   {\cal M}^{(1)}_{s, H^\pm} &=
   Z^2 \begin{bmatrix}
    0 & 0 & 0 & 0 \\[-.3cm]
    0 & 1 & 1 & 0 \\[-.3cm]
    0 & 1 &1 & 0 \\[-.3cm]
    0 & 0 & 0 & 0 \\
   \end{bmatrix} \ ;
   \\
   \label{eq:m2u}
   {\cal M}^{(1)}_{u,G^\pm} &=
   Z^2 \begin{bmatrix}
    1 & 0 & 0 & 0 \\[-.3cm]
    0 & 0 & 1 & 0 \\[-.3cm]
    0 & 1 & 0 & 0 \\[-.3cm]
    0 & 0 & 0 & 1 \\
   \end{bmatrix}\ ,
   &
   {\cal M}^{(1)}_{u,H^\pm} &=
   Z^2 
   \begin{bmatrix}
    0 & 0 & 0 & \pm 1 \\[-.3cm]
    0 & 1 & 0 & 0 \\[-.3cm]
    0 & 0 &1 & 0 \\[-.3cm]
   \pm 1 & 0 & 0 & 0 \\
   \end{bmatrix}\ ;
   \\  \label{eq:mtha} 
   {\cal M}^{(1)}_{t,h} &= Z^2
   \begin{bmatrix}
    1 & 0 & 0 & 0 \\[-.3cm]
    0 & 1 & 0 & 0 \\[-.3cm]
    0 & 0 & 1 & 0 \\[-.3cm]
    0 & 0 & 0 & 1 \\
   \end{bmatrix}\ ,
   &
   {\cal M}^{(1)}_{t,G^0} &= 
   \begin{bmatrix}
    0 & 0 & 0 & 0 \\[-.3cm]
    0 & 0 & 0 & 0 \\[-.3cm]
    0 & 0 & 0 & 0 \\[-.3cm]
    0 & 0 & 0 & 0 \\
   \end{bmatrix} \ ,
   \\ \nonumber
   {\cal M}^{(1)}_{t,H/A} &= Z^2
   \begin{bmatrix}
    0 & 0 & 0 & \pm 1\\[-.3cm]
    0 & 0 & 1 & 0 \\[-.3cm]
    0 &1 & 0 & 0 \\[-.3cm]
   \pm 1 & 0 & 0 & 0 \\
   \end{bmatrix}\ ,
  &
  {\cal M}^{(1)}_{t,A/H} &=
   \begin{bmatrix}
    0 & 0 & 0 & 0 \\[-.3cm]
    0 & 0 & 0 & 0 \\[-.3cm]
    0 & 0 & 0 & 0 \\[-.3cm]
    0 & 0 & 0 & 0 \\
   \end{bmatrix} \ .
\end{align}
In the last two matrices in Eq.~(\ref{eq:mtha}), the $+1$ in the off-diagonal entries is for the $H$-mediated  diagram in the $Z_i=Z_5$ solution, while the $A$-mediated  diagram vanishes. On the other hand, in the $Z_i=-Z_5$ solution the $H$-mediated diagram vanishes and the $A$-mediated diagram takes $-1$ in the (1,4) and (4,1) entries. 

In each $s/t/u$-channel, there are two diagrams mediated by different particles and the flavor subamplitudes are therefore weighted by different propagators $P_{r,s}$. The only way entanglement can be maximized in each channel is if the two particles have degenerate mass and the flavor subamplitudes can be added with equal weight. For instance, the $s/u$-channel amplitudes will maximize entanglement only if $m_{G^\pm}=m_{H^\pm}=0$ such that  the flavor subamplitudes in Eqs.~(\ref{eq:m2s}) and (\ref{eq:m2u}) conspire to produce a maximally entangled state. Similarly, the $t$-channel amplitude maximizes entanglement only when $m_h=m_{H}$ (for $Z_i=Z_5$) or $m_h=m_{A}$ (for $Z_i=-Z_5$), in which cases the propagators can again be factored out. Summing over the quartic and $s/t/u$-channel contributions, the full amplitude now has the flavor structure,
\be
{\cal M}^{(1)} \propto  \begin{bmatrix}
    1 & 0 & 0 & \pm 1 \\[-.3cm]
    0 & 1 & 1 & 0 \\[-.3cm]
    0 & 1 & 1 & 0 \\[-.3cm]
    \pm 1 & 0 & 0 & 1 
   \end{bmatrix} \ , \label{eq::cn4sym}
\ee
which is the same flavor structure produced by the quartic couplings in the unbroken phase.

The requirement of $m_{G^\pm}=m_{H^\pm}=0$ arises from the $s/u$-channel and leads to 
\be
\label{eq:yyy}
Y\equiv Y_1=Y_2= -\frac12 Zv^2 \ ,
\ee
as can be seen from Eq.~(\ref{eq:masq}). The quadratic terms in the scalar potential now are 
\be 
V^{(2)} = Y\left(\Phi^\dagger_1 \Phi_1+\Phi^\dagger_2 \Phi_2\right) \ ,
\ee
which has the maximal $U(4)$ symmetry. Together with the quartic couplings in Eqs.~(\ref{eq:vp4}) and (\ref{eq:vm4}), entanglement maximization in the broken phase of 2HDM now leads to enhanced $U(2)_{\sf{S}}\times U(2)_{\sf{A}}$ or $U(2)_{\sf{L}}\times U(2)_{\sf{R}}$ symmetry. It should be emphasized that the symmetry is respected by both the quartic and quadratic terms in the scalar potential, unlike in the unbroken phase where the symmetry was broken softly by the mass terms.

Using Eq.~(\ref{eq:yyy}) in the other mass matrices in Eqs.~(\ref{eq:mcpe}) and (\ref{eq:mcpo}), we arrive at $m_h=m_H$ if $Z_i=Z_5$ or $m_h=m_A$ if $Z_i=-Z_5$, exactly what is required to gives rise to maximal entanglement in the $t$-channel. In the end, the spectrum consists of 6 massless Nambu-Goldstone modes and two degenerate massive scalars. To understand this pattern, it is easiest to use the basis of $\Phi_{S/A}$ and $\Phi_{L/R}$ defined in Eqs.~(\ref{eq:phias}) and (\ref{eq:philr}). In these bases the scalar potential decomposes into two independent sectors:
\be
\label{eq:twosu2}
V= \sum_{\substack{\Gamma=\sf{S,A}\\ {\rm or}\ \sf{L, R}}} V_\Gamma \ ,\qquad 
V_{\Gamma} = Y\, \Phi^\dagger_\Gamma \Phi_\Gamma + Z \left(\Phi^\dagger_\Gamma \Phi_\Gamma \right)^2 \ .
\ee
The two $U(2)$'s act independently on these two sectors.  The spontaneous symmetry breaking in the Higgs basis comes from 
$\langle H_1^\dagger H_1 \rangle = v^2/2$ and $\langle H_2^\dagger H_2\rangle =0$, which in the new bases translate into
\be
\langle \Phi_{\sf{S}}^\dagger \Phi_{\sf{S}} \rangle =\langle \Phi_{\sf{A}}^\dagger \Phi_{\sf{A}} \rangle = \langle \Phi_{\sf{L}}^\dagger \Phi_{\sf{L}}  \rangle = \langle \Phi_{\sf{R}}^\dagger \Phi_{\sf{R}} \rangle =\frac{1}{2} \langle H_1^\dagger H_1  \rangle =\frac{v^2}{4}\ ,
\ee
\be
V= \sum_{\substack{\Gamma=\sf{S,A}\\ {\rm or}\ \sf{L, R}}}  Z \left( \Phi^\dagger_\Gamma \Phi_\Gamma - \frac{v^2}{4} \right)^2 \ .
\ee 
It follows that  each of the $U(2)$'s is broken spontaneously to $U(1)$'s, $U(2)_{\sf{S/L}}\times U(2)_{\sf{A/R}} \to U(1)_{\sf{S/L}}\times U(1)_{\sf{A/R}}$, giving rise to six massless Goldstone modes in total and two degenerate massive scalars.

In the Appendix~\ref{sect:app} we go beyond the charged-neutral scattering and list the amplitudes for the other scattering channels. One sees that the same enhanced symmetries also maximize entanglement in all other channels.

 Let us stress for clarity that when  $Z_6 = Z_7 = 0$ and $Y_3 = 0$, as is required for maximal entanglement in the Higgs basis, the phase of $Z_5$ in the potential can be modified by  a phase rotation of the Higgs field $H_2$. Therefore the two cases we are studying with $Z_5 >0$ and $Z_5 <0$ can be mapped into each other upon a phase rotation $H_2 \to i H_2$, which amounts to rotating the $|1 \rangle$ in the computational basis by an
imaginary factor (see Eq.~(\ref{eq:imqubit})). This same imaginary factor converts the $\mathsf{S,A}$ basis in flavor space into the $\mathsf{L,R}$ one, as is apparent from Eqs.~(\ref{eq:phias}) and (\ref{eq:philr}).

\section{Gauge Interactions and Custodial Invariance}
\label{sect:5}

Having demonstrated that entanglement maximization in 2HDM leads to enhanced $U(2)\times U(2)$ symmetries, next we consider whether maximal entanglement can be maintained when  the enhanced global symmetry is promoted to a gauge symmetry. In the following we will gauge the $SU(2)\times SU(2)$ subgroup and study whether entanglement maximization can still be achieved.  

It is worth recalling that the ``vector'' and ``axial'' generators of the $SU(2)\times SU(2)$ group are given by Eqs.~(\ref{eq:vplus}) and (\ref{eq:vminus}), 
\be
T_{\mathsf{v}}^a = \openone\otimes \frac{\sigma^a}2 \ , \qquad 
T_{\mathsf{a}}^a = \tau^i \otimes \frac{\sigma^a}2 \ , \quad a=1,2,3\ ,
\ee
where $i=1,2$ for $Z_i=\pm Z_5$, respectively. Recall $T_{\mathsf{v}}^a$ generates the electroweak $SU(2)_L$. The product group structure is transparent when we go to the basis
\be
T^a_{\mathsf{S/L}} = \frac{\openone+\tau^i}2\otimes \frac{\sigma^a}2 \ ,\qquad 
T^a_{\mathsf{A/R}} = \frac{\openone-\tau^i}2\otimes \frac{\sigma^a}2 \ , \quad i=1,2\ ,
\ee
where $T^a_{\mathsf{v}}= T^a_{\mathsf{S/L}}+T^a_{\mathsf{A/R}}$ and $T^a_{\mathsf{a}}= T^a_{\mathsf{S/L}}-T^a_{\mathsf{A/R}}$.
The gauge fields are $A_{\Gamma}^\mu = A_{{\Gamma}}^{\mu a}\ T_{\Gamma}^a$, $\Gamma=\{\mathsf{v,a,S,A,L,R}\}$. 
In the four-component notation $\Phi=(\Phi_1,\Phi_2)^T$, the covariant derivative is given by
\be
D^\mu \Phi = \left(\partial^\mu + ig_{\mathsf{v}} A_{\mathsf{v}}^{\mu} + ig_{\mathsf{a}} A_{\mathsf{a}}^{\mu}\right)\Phi\ ,
\ee
where 
\bea
\label{eq:vgauge}
g_{\mathsf{v}}A_{\mathsf{v}}^{\mu\,a} &=& \frac12 \left(g_{\mathsf{S}}A_{\mathsf{S}}^{\mu\,a}+g_{\mathsf{A}}A_{\mathsf{A}}^{\mu\,a}\right)\quad {\rm or}\quad \frac12 \left(g_{\mathsf{L}}A_{\mathsf{L}}^{\mu\,a}+g_{\mathsf{R}}A_{\mathsf{R}}^{\mu\,a}\right) \ ,\\
\label{eq:agauge}
g_{\mathsf{a}}A_{\mathsf{a}}^{\mu\,a} &=& \frac12 \left(g_{\mathsf{S}}A_{\mathsf{S}}^{\mu\,a}-g_{\mathsf{A}}A_{\mathsf{A}}^{\mu\,a}\right)\quad {\rm or}\quad \frac12 \left(g_{\mathsf{L}}A_{\mathsf{L}}^{\mu\,a}-g_{\mathsf{R}}A_{\mathsf{R}}^{\mu\,a}\right)\ .
\eea
The kinetic terms for   the scalar doublets and the gauge fields are
\bea
\label{eq:scalarke}
{\cal L}_{\rm KE} &=& \left(D^\mu\Phi\right)^\dagger\left(D_\mu\Phi\right) = \sum_{\substack{\Gamma=\mathsf{S,A}\\ {\rm or}\ \mathsf{L, R}}} \left(D^\mu\Phi_{\Gamma}\right)^\dagger\left(D_\mu\Phi_{\Gamma}\right)\ ,\\
{\cal L}_{\rm YM} &=& \sum_{\substack{\Gamma=\mathsf{S,A}\\ {\rm or}\ \mathsf{L, R}}} -\frac14 \left(F_{\Gamma\,\mu\nu} F_\Gamma^{\mu\nu}\right) \ ,\qquad F_\Gamma^{\mu\nu}= \partial^\mu A_\Gamma^\nu - \partial^\nu A_\Gamma^\mu +i g_\Gamma [A_\Gamma^\mu,A_\Gamma^\nu] \ ,
\eea
where $D^\mu\Phi_\Gamma= \partial^\mu \Phi_\Gamma+ ig_\Gamma A_\Gamma^\mu \Phi_\Gamma$ is the covariant derivative in the ``two-component'' notation since $\Phi_\Gamma$, $\Gamma=\{\sf{S,A,L,R}\}$ are two-compnent doublet scalars. Moreover, notice that the $\sf{S/L}$ and $\sf{A/R}$ sectors decouple from each other because of the product group structure.

To compute the 2-to-2 scattering when turning on the gauge fields, it is necessary to choose a gauge. For spontaneously broken gauge theories,   it is popular to choose the unitary gauge at the tree-level, where the eaten Goldstone components are set to zero and the gauge fields contain both the longitudinal and transverse components. For our purpose it is slightly more convenient to choose a renormalizable $R_\xi$ gauge, where we retain the Goldstone degrees of freedom. In particular, we will choose the Lorentz gauge, $\xi=0$, where the Goldstone  modes remain massless and have  the same couplings as in
the ungauged theory, while the gauge boson propagator
is purely transverse \cite{Peskin:1995ev}:
\be
\langle W^a_\mu(p) W^b_\nu(-p)\rangle = \frac{-i\delta^{ab}}{p^2-m_W^2} \left(g_{\mu\nu}-\frac{p_\mu p_\nu}{p^2}\right) \ , \qquad
\langle G^{a}(p) G^{b}(-p)\rangle = \frac{i\delta^{ab}}{p^2}\ .
\ee
In the $\xi=0$ gauge, and more generally in any renormalizable gauge, one can still talk about, for instance, $H^+_a H^0_b \to H^+_c H^0_d$ scattering even in the presence of gauge interactions. We will briefly comment on the physical picture in the unitary gauge later. Since the Goldstone modes have the same couplings as in the ungauged theory, all the calculations in Sect.~\ref{sect:3} remain valid and we only need to consider new contributions mediated by the gauge bosons, which result from the three-point couplings in the kinetic term in Eq.~(\ref{eq:scalarke}).

Since we are mainly interested in the flavor quantum number of the outgoing particles for those diagrams mediated by the gauge fields, the analysis proceeds in a way similar to that in  Eqs.~(\ref{eq:XX1}) -- (\ref{eq:YY1-}), which directly leads to the flavor structures in Eqs.~(\ref{eq:mm1}) and (\ref{eq:mm2}). In essence, because $A_\Gamma^\mu$ only mediates scatterings between $\Phi_\Gamma$ particles, the contributions from gauge-mediated diagrams are the sum of two decoupled sectors. Using $H^+H^0$ as an example,  requiring the amplitudes from gauge-mediated $t$-channel diagram in the $Z_i=Z_5$ solution to produce the same maximally entangled flavor states as in the ungauged case, 
\bea
\label{eq:mta}
{\cal M}_{t,A^\mu}^{(1)}&=& P_{t,\sf{S}}\, f^{(1)}_{t,\sf{S}} \ \frac{\openone+\tau^1}2\otimes \frac{\openone+\tau^1}2+P_{t,\sf{A}}\,f^{(1)}_{t,\sf{A}}\ \frac{\openone-\tau^1}2\otimes \frac{\openone-\tau^1}2 \nonumber \\
&\propto& P_{t}\, f^{(1)}_{t}\ \frac{\openone\otimes\openone+\tau^1 \otimes\tau^1}2 \ ,
\eea
where $P_{t,\sf{S/A}}$ is the propagator for $A^\mu_{\sf{S/A}}$ and $f^{(2)}_{t,\sf{S/A}}$ contains the channel dependent kinematic factors as well as the gauge couplings $g_{\sf{S}}$ and $g_{\sf{A}}$. Its expression (including the additional channels) reads:
\begin{equation}\label{eq:sagauge}
f_r^\Gamma = \mathcal{C}\times\frac{g_\Gamma^2}{2}\times
\begin{cases}
       (p_a - p_b)^\mu(p_c - p_d)^\nu,\quad r=s \\
       (p_a - p_c)^\mu(p_b - p_d)^\nu,\quad r=t \\
       (p_a - p_d)^\mu(p_b - p_c)^\nu,\quad r=u
   \end{cases}\,; \qquad \mathcal{C}=
   \begin{cases}
      1, \quad A^{1,2}_\Gamma \\
      \frac{1}{2}, \quad A^3_\Gamma 
   \end{cases},
\end{equation}
where $\{a,b,c,d\}$ label the entries of the matrix amplitude in Eq.~(\ref{eq:sagauge}).
One must have $ P_{t,\sf{S}} f^{(1)}_{t,\sf{S}}= P_{t,\sf{A}} f^{(1)}_{t,\sf{A}}=P_{t} f^{(1)}_{t}$, which need to hold not only for the $s/t/u$-channel diagrams, but also for every scattering channel in Eqs.~(\ref{eq::polish}) -- (\ref{eq::polish1}), thereby implying $g_{\sf{S}}=g_{\sf{A}}$ and degenerate mass for the gauge bosons in the $\sf{S}$- and $\sf{A}$-sector. These relations are guaranteed by the existence of a discrete $\mathrm{Z}_2$ symmetry interchanging the two sectors.   This $\mathrm{Z}_2$ symmetry is defined so that
\begin{equation}
H_1 \to H_1, \;\;\; H_2 \to -H_2, \quad  \Phi_{A} \leftrightarrow \Phi_{S}, \;\;\; A^\mu_{A} \leftrightarrow A^\mu_{S} .
\end{equation}
Observe that the Higgs bosons in the computational basis are associated with the $\pm 1$ Higgs eigenstates of the discrete
$\mathrm{Z}_2$ symmetry.

The analysis for the $Z_i=-Z_5$ solution is similar to those surrounding Eqs.~(\ref{eq:YY1+}) and (\ref{eq:YY1-}). For $I=1,2$ scattering channel, the gauge-mediated amplitudes are the same as in Eq.~(\ref{eq:mta}), with the $\sf{S}$- and $\sf{A}$-sector replaced by $\sf{L}$- and $\sf{R}$-sector. For $I=3,4,5$ channels, the gauge-mediated amplitudes decompose in a way similar to Eq.~(\ref{eq:YY1-}),
\bea
\label{eq:mtaa}
{\cal M}_{r,A^\mu}^{(I)}&=& P_{r,\sf{L}}\, f^{(I)}_{r,\sf{L}} \ \frac{\openone+\tau^2}2\otimes \frac{\openone-\tau^2}2+P_{r,\sf{R}}\,f^{(I)}_{r,\sf{R}}\ \frac{\openone-\tau^2}2\otimes \frac{\openone+\tau^2}2 \nonumber \\
&\propto& P_{r}\, f^{(I)}_{r}\ \frac{\openone\otimes\openone-\tau^2\otimes\tau^2}2 \ , \qquad r=s, t, u\ ;\ \ I=3,4,5 \ .
\eea
Again we obtain $g_{\sf{L}}=g_{\sf{R}}$ and degenerate mass in the $\sf{L}$- and $\sf{R}$-sector. We also see the emergence of a discrete $\mathrm{Z}_2$ symmetry: 
$
\Phi_{L} \leftrightarrow \Phi_{R}, \, A^\mu_{L} \leftrightarrow A^\mu_{R}$.

\begin{figure}
    \centering
    \includegraphics[width=1\linewidth]{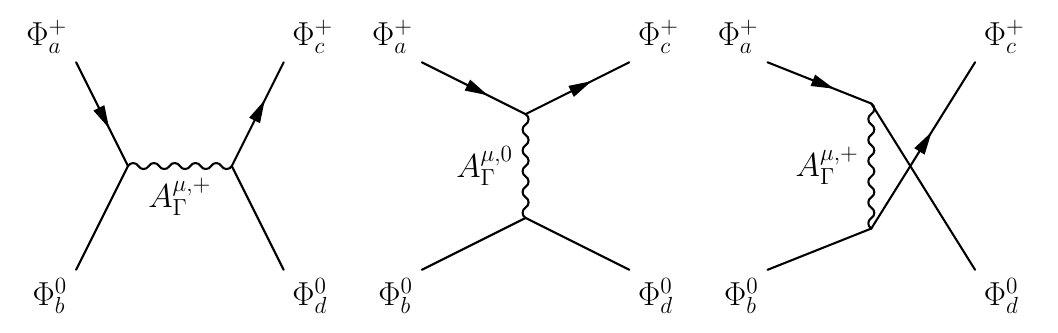}
    \caption{Gauge-mediated Feynman diagrams for $H^+H^0\to H^+H^0$ in the $\sf{S/A}$ basis (see Eq.~(\ref{eq:phias},\ref{eq:philr})), where basis elements are labeled with $\{a,b,c,d\}\in\{\sf{S,A}\}$, and $A_\Gamma^{\mu,+} = \frac{A_\Gamma^{\mu,1} + i A_\Gamma^{\mu,2}}{\sqrt{2}}$ with $\Gamma \in \{\sf{S,A}\}$.}
    \label{fig::gauge}
\end{figure}

For the sake of clarity, we provide here the matrix elements for the $H^+ H^0 \to H^+ H^0$ scattering. As discussed above, given the manifest product structure, it is easier to perform the calculations in the $\sf{S/A}$ or $\sf{L/R}$ basis. We focus on the first one (therefore on the $Z_i=Z_5$ symmetry), as the computation is similar for the latter. The flavor structure is the same for each channel (see Fig.~\ref{fig::gauge}) and reads:

\begin{equation}
   {\cal M}^{(1)}_{r,A^\mu} = -  P_{r,\sf{S}} f_{r,\sf{S}}   
   \begin{bmatrix}
    1 & 0 & 0 & 0 \\[-.3cm]
    0 & 0 & 0 & 0 \\[-.3cm]
    0 & 0 & 0 & 0 \\[-.3cm]
    0 & 0 & 0 & 0  
   \end{bmatrix} -  P_{r,\sf{A}} f_{r,\sf{A}}   
   \begin{bmatrix}
    0 & 0 & 0 & 0 \\[-.3cm]
    0 & 0 & 0 & 0 \\[-.3cm]
    0 & 0 & 0 & 0 \\[-.3cm]
    0 & 0 & 0 & 1  
   \end{bmatrix} \; , \; r=(s,t,u) 
\end{equation}

Imposing $\Delta=1$ on each channel separately, yields maximal entanglement and enforces the gauge couplings to be the same ($g_{\sf{S}}=g_{\sf{A}}=g$) together with degenerate gauge boson masses in the $\sf{S/A}$-sectors. Under these conditions, the matrix elements for each channel in the original computational basis are:
\begin{align} \label{eq::rotation}
   {\cal M}^{(1)}_{r,A^\mu} &= - \frac{g^2}{2} P_{r}f_{r}   
   \begin{bmatrix}
    1 & 0 & 0 & 0 \\[-.3cm]
    0 & 0 & 0 & 0 \\[-.3cm]
    0 & 0 & 0 & 0 \\[-.3cm]
    0 & 0 & 0 & 1  
   \end{bmatrix} \;
   \xrightarrow[\text{computational basis}]{\text{flavor rotation to}}
   \; - \frac{g^2}{4} P_{r}f_{r}
   \begin{bmatrix}
    1 & 0 & 0 & 1 \\[-.3cm]
    0 & 1 & 1 & 0 \\[-.3cm]
    0 & 1 & 1 & 0 \\[-.3cm]
    1 & 0 & 0 & 1 
   \end{bmatrix} \; , \; r=(s,t,u),
\end{align}
With $P_r = P_{r,\sf{S}} = P_{r,\sf{A}}$ and $f_r = f_{r,\sf{S}} = f_{r,\sf{A}}$. Notice that we recover the structure of the quartic interactions (see Eq.~(\ref{eq::cn4sym})) and that the gauge-mediated amplitudes convert basis elements of the original computational basis into Bell states.

It is informative to consider the same scattering processes in the unitary gauge where the Goldstone modes are eliminated completely and become the longitudinal components of the now massive gauge bosons. In the scalar sector the remaining degrees of freedom are the massive scalars $h/H$ or $h/A$, depending on whether $Z_i=Z_5$ or $Z_i=-Z_5$. Recall that we define the computational basis using the flavor of the doublet scalars $|0\rangle=|H_1\rangle$ and $|1\rangle=|H_2\rangle$. While in the $R_\xi$ gauge all the Goldstone modes remain in the theory and diagrammatically one can still talk about, say, $H_a^+H_b^- \to H_c^+H_d^-$, it is less clear how to define such a scattering process when $H^\pm$ is removed from the theory in the unitary gauge. In this regard, it is helpful to identify that $H^\pm_1$ and $H^\pm_2$ are the longitudinal components of the vector and axial gauge fields, respectively, defined in Eqs.~(\ref{eq:vgauge}) and (\ref{eq:agauge}). This can be seen from the fact that $H_1=(\Phi_{\sf{S/L}}+\Phi_{\sf{A/R}})/\sqrt{2}$ and $H_2=(\Phi_{\sf{S/L}}-\Phi_{\sf{A/R}})/\sqrt{2}$. The observation suggests that, for example, the scattering process corresponding to $H_1^+H_2^- \to H_1^+H_2^-$ in the unitary gauge is $A_{\sf{v}}^{(L)}A_{\sf{a}}^{(L)} \to A_{\sf{v}}^{(L)}A_{\sf{a}}^{(L)} $, where $A_{\sf{v,a}}^{(L)}$ refers to the longitudinal component of $A_{\sf{v,a}}^\mu$. In other words, the massive gauge fields  now constitute a ``qubit'' in the space of vector and axial gauge fields, $|A_{\sf{v}}\rangle = |0\rangle$ and $|A_{\sf{a}}\rangle=|1\rangle$. The computational basis for the 2-to-2 scattering of vector bosons can be defined accordingly: $\{|\sf{v}\sf{v}\rangle, |\sf{v}\sf{a}\rangle, |\sf{a}\sf{v}\rangle,|\sf{a}\sf{a}\rangle\}$. It is with respect to this basis that the entanglement in the vector-boson scattering is maximized in the unitary gauge.

It should also be clear that the above arguments about entanglement maximization are still valid if the remaining $U(1)\times U(1)$ are gauged. Similar to the case of gauging $SU(2)\times SU(2)$, maximal entanglement requires the appearance of a discrete $\mathrm{Z}_2$ symmetry enforcing the equality of the $U(1)$ gauge couplings. Therefore, if the full SM gauge group were gauged, two identical copies of the SM gauge bosons would be required for maximal entanglement.
In the Standard Model with only one Higgs doublet $\Phi=(\phi_1+i\phi_2, \phi_3+i\phi_4)^T$, it is well-known that the scalar sector has an accidental $SO(4)$ symmetry protecting the  electroweak oblique $T$-parameter from deviating from unity: $\Delta T=0$ \cite{Peskin:1990zt}. This ``custodial symmetry'' stems from the fact that a scalar potential of the form $\mu^2\, \Phi^\dagger\Phi+\lambda (\Phi^\dagger\Phi)^2$ can be rewritten as $\mu^2\,\vec{\Phi}\cdot\vec{\Phi}+\lambda (\vec{\Phi}\cdot\vec{\Phi})^2$, which is then invariant under an $O(4)$ rotation of $\vec{\Phi}=(\phi_1,\phi_2,\phi_3,\phi_4)^T$. In our setup, it is evident from Eq.~(\ref{eq:twosu2}) the scalar potential in the 2HDM actually has two copies of custodial $SO(4)$ invariance, with one of them becoming the custodial $SO(4)$ in the Standard Model. Therefore we expect vanishing scalar contributions to the electroweak  $T$-parameter. After including the fermions the custodial invariance is violated at the loop level, just as in the SM. This becomes quite clear in the mirror fermion construction described in the next section, in which the SM structure is preserved.

\section{Loop Corrections and the Inclusion of Fermions}
\label{sect:loop}

In the previous sections, we have studied the requirement to obtain maximal flavor entanglement in tree-level scattering amplitudes  of Higgs bosons, starting with separable states in the computational basis. The resulting theory has an extended $SU(2) \times U(1) \times SU(2) \times U(1)$ gauge symmetry, with an additional ${\mathrm Z}_2$ discrete symmetry relating the two gauge and Higgs sectors. Although the study was based on tree-level amplitudes, the emergent symmetry implies that the relation between the quartic and gauge couplings will be stable under renormalization group evolution at any order in perturbation theory. 

The inclusion of fermions will affect the scattering amplitudes at the loop level. At the one-loop level, their contribution will be given by box diagrams, and therefore proportional to four Yukawa coupling factors. In the case of $\Phi_{S,A}$ The Yukawa couplings of the SM fermions are given by
\begin{equation}
{\mathcal{L}} = y_S \Psi_L \Phi_S \psi_R
+ y_A \Psi_L \Phi_A \psi_R + h.c.,
\end{equation}
where $\Psi_L$ and $\psi_R$ represent a given SM quark or lepton doublet and singlet, respectively.  It is clear at this point that the only way of preserving the discrete $\mathrm{Z}_2$ symmetry is by establishing a relationship between the $y_S$ and $y_A$ couplings. However, the $\mathrm{Z}_2$ symmetry is not enough to ensure maximal entanglement. 
It is also important that the final states
contain only pairs of $\Phi_S$ and $\Phi_A$
states, without any mixed $\Phi_S$ $\Phi_A$ state. Observe that at every vertex of the box diagram, one would be able to produce $\Phi_S$ or $\Phi_A$ states with an amplitude proportional to $y_S$ and $y_A$, respectively. Therefore, the final states will include $\Phi_{S} \Phi_S$ and $\Phi_A \Phi_A$ states, with amplitudes proportional to $y_S^2$ and $y_A^2$, as well as mixed $\Phi_S \Phi_A$ states with amplitudes proportional to $y_S y_A$.

\begin{figure}
    \centering
    \includegraphics[width=1\linewidth]{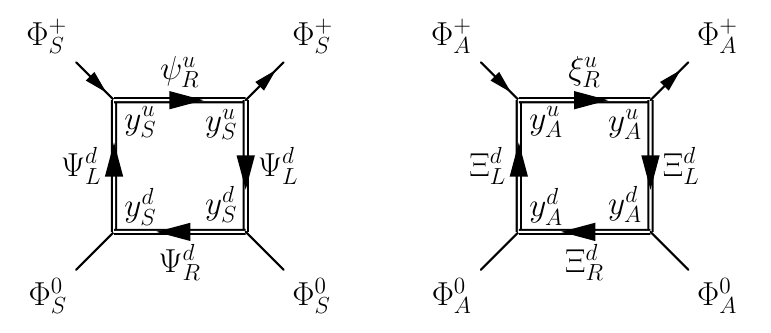}
    \caption{Representative Feynman diagrams induced via a fermion loop for $H^+H^0\to H^+H^0$ scattering in the $\sf{S/A}$ basis (see Eq.~(\ref{eq:phias},\ref{eq:philr})). Only the represented flavor configurations are allowed.} 
    \label{fig::fermion}
\end{figure}

It is clear from the above discussion that it is not possible to include just the SM fermions and keep the conditions leading to maximal entanglement. In order to recover maximal entanglement, new mirror fermions must be included. One possibility is to associate the $\Phi_S$ and $A_\mu^S$ fields with the SM Higgs and gauge fields, being coupled to the SM fermions which will carry no charge under the non-standard gauge group associated with the fields $A_\mu^A$. Similarly, the mirror fermions $\Xi, \xi$ will only couple to $\Phi_A$ and $A_{\mu}^A$ and will carry no charges under the SM gauge group. The Yukawa terms read

\begin{equation}
    \mathcal{L}= y^u_S \bar\Psi_L \tilde{\Phi}_S \psi^u_R + y^d_S \bar\Psi_L \Phi_S \psi^d_R
+ y^u_A \bar\Xi_L \tilde{\Phi}_A \xi^u_R + y^d_A \bar\Xi_L \Phi_A \xi^d_R + h.c.
\end{equation}
where $\tilde{\Phi}_{\sf{S}/\sf{A}} = i\tau^2 \Phi_{\sf{S}/\sf{A}}$.
It is now clear that, under these conditions, considering the gauge and Yukawa couplings of the SM and mirror fermions to be the same, 
\begin{equation}
y^u_S = y^u_A, \;\;\;\;\;\;\;\;  y^d_S = y^d_A,
\end{equation}
one keeps the $\mathrm{Z}_2$ symmetry by generalizing it to the fermion sector,
\begin{equation}
    \Psi_L \to \Xi_L,  \;\;\; \psi^u_R \to \xi^u_R , \;\;\; \psi^d_R \to \xi^d_R ,
\end{equation}
and one recovers  the production of only pairs of $\Phi_S$ and $\Phi_A$ bosons, associated with loops of SM and mirror fermions, respectively (see Fig.~\ref{fig::fermion}). For the sake of clarity let us also write explicitly the loop amplitudes for our usual proxy scattering $H^+ H^0 \to H^+ H^0$ in the $\sf{S}/\sf{A}$ basis:

\begin{equation}
   {\cal M}^{(1)}_{F} = L_{\sf{S}}(y^u_{\sf{S}} y^d_{\sf{S}})^2   
   \begin{bmatrix}
    1 & 0 & 0 & 0 \\[-.3cm]
    0 & 0 & 0 & 0 \\[-.3cm]
    0 & 0 & 0 & 0 \\[-.3cm]
    0 & 0 & 0 & 0  
   \end{bmatrix} +  L_{\sf{A}}(y^u_{\sf{A}} y^d_{\sf{A}})^2   
   \begin{bmatrix}
    0 & 0 & 0 & 0 \\[-.3cm]
    0 & 0 & 0 & 0 \\[-.3cm]
    0 & 0 & 0 & 0 \\[-.3cm]
    0 & 0 & 0 & 1  
   \end{bmatrix}
\end{equation}
where $L_{\sf{S/A}}$ represents the loop factors. The same argument as for the gauge mediated diagrams applies also here, requring the same Yukawa couplings and the same fermion masses for the two $\sf{S/A}$ sectors. Employing a rotation to the computation basis, one recover the same flavor structure as in Eq.~(\ref{eq::rotation}), where the maximization of the entanglement is manifest.

Hence, maximal entanglement demands not only the duplication of symmetries, with a degenerate gauge and Higgs boson spectrum, but also the introduction of a whole dark, mirror fermion sector with equal mass and flavor structure as the SM one, which is decoupled from the SM interactions. However, the absence of interactions between the SM and the mirror world would prevent us to conduct  an experiment observing the flavor entanglement, as there is no way to produce the initial states in the computational basis, which are superpositions of the SM and the mirror states. 

To avoid this problem, we could identify the  standard model fermions with the $+1$ eigenstates of the $\mathrm{Z}_2$ symmetry and mirror fermions with the $-1$ ones, namely the orthogonal linear combinations of $\Psi, \psi$ and $\Xi,\xi$, for example
\begin{equation}
    \Psi_{SM}=\frac 1 {\sqrt{2}}( \Psi+\Xi), \;\;\; \Psi_{Mirror}=\frac 1 {\sqrt{2}}( \Psi-\Xi),
\end{equation} 
and similarly for $\psi, \xi$. The SM gauge and Higgs bosons would also be identified with the $+1$ eigenstates, while the non-standard bosons will be associated with the $-1$ ones. In this case the standard and non-standard sectors interact with each other, with couplings that preserve the $\mathrm{Z}_2$ symmetry. This would enable the production of computational basis states, which become maximally entangled after the scattering. However, such a model  presents phenomenological problems, since the SM Higgs and gauge bosons are coupled to two copies of fermions of the same mass. Hence this is not a phenomenologically realistic scenario.

Let us just finish this section by stressing that all that we have said about the $\Phi_{S,A}$ case applies also to the $\Phi_{L,R}$ case. Hence, in both cases the requirement of maximal flavor entanglement in the final states of the scattering of Higgs states in the computational basis leads to the presence of a duplication of the symmetries and a new dark mirror world, decoupled from the SM interactions. 

 Let us finish by stressing that the theory at the weak scale may be an effective low energy theory coming from ultraviolet interactions that connect the two sectors, and are suppressed at weak scale energies. The computational basis states may be produced by these suppressed interactions. The question of entanglement in the high energy theory would depend on the extra degrees of freedom at the high energy scale. In addition, the standard and mirror sectors will couple to gravity and hence, although the low energy model presents no phenomenological problems, it will lead to cosmology questions that need to be addressed, something that also depends on the ultraviolet completion and is hence beyond the scope of this article.

\section{Conclusion}
\label{sect:con}

In this work we present an example of enhanced symmetries arising from entanglement maximization. We consider a general CP-conserving 2HDM and require the flavor entanglement in the outgoing states of 2-to-2 scattering of Higgs bosons to be maximal when the incoming flavor quantum numbers are unentangled. Specifically, we compute the outgoing entanglement in flavor quantum numbers as a function of coupling constants in the scalar potential and demand the entanglement be maximal when the  flavor of the two incoming particles is a basis vector in the computational basis. Contrary to some earlier works, we do not restrict ourselves to special corners of phase space, but instead require the outgoing flavors to be maximally entangled over the entire kinematic phase space.

We start by considering the 2HDM with unbroken  electroweak $SU(2)_L\times U(1)_Y$ symmetry. Turning off  gauge interactions, the 2-to-2 scattering is mediated only by 4-point contact interactions. We find three families of solutions, each containing two solutions with opposite sign choices among otherwise equal sets of quartic couplings. These solutions give rise to maximal entanglement over the full product basis in the computational basis. However, only one family of solutions, $\lambda_1=\lambda_2=\lambda_3=\lambda_4=\pm \lambda_5$ and $\lambda_6=\lambda_7=0$,  results in a scalar potential that is bounded from below, and we therefore focus on it. In this case, both sign choices enhance the global symmetry from $U(2)$ to  $U(2)\times U(2)$ among the quartic couplings. The quadratic terms in the potential are not constrained by entanglement maximization in the unbroken phase. As a consequence, the $U(2)\times U(2)$ symmetry is in general broken softly by the mass terms. Moreover, the presence of the enhanced  $U(2)\times U(2)$ symmetry does not depend on the electric charge of the Higgs bosons participating in the scattering process. 

Next we study the 2HDM in the spontaneously broken phase in the Higgs basis, where only one of the Higgs doublets  acquires the full VEV. In the broken phase there are three-point vertices generated by the VEV and the 2-to-2 scattering now contains contributions from $s/t/u$-channels, in addition to the 4-point contact interactions. Since we demand entanglement maximization over the entire kinematic phase space, this is possible only when the $s/t/u$-channel contributions produce the same flavor structure as in the contact interactions, which in turn requires degenerate masses among the scalars mediating diagrams in the same channel. These considerations constrain the quadratic terms in the scalar potential and the enhanced $U(2)\times U(2)$ now becomes exact and is spontaneously broken to $U(1)\times U(1)$, giving rise to 6 massless Goldstone bosons and two degenerate massive scalars.

The massless Goldstones can be lifted by gauging an $SU(2)\times U(1) \times SU(2) \times U(1)$ subgroup. Even though the gauge bosons now induce additional contributions to the 2-to-2 scattering process, we show that maximal entanglement can be preserved if a discrete $\mathrm{Z}_2$ symmetry is introduced which interchanges the two gauge sectors and ensures equal gauge coupling strengths. All massless Goldstones now become the longitudinal components of the massive gauge bosons. In the unitary gauge, the scattering of Goldstone bosons corresponds to the scattering of the longitudinal component of the massive gauge bosons.

The introduction of SM fermions in the theory affects the scattering amplitudes at the loop level and makes it impossible to preserve the maximal entanglement conditions.  Maximal entanglement can only be ensured in this case by introducing a new set of Dark Mirror fermions, which couple to only one of the Higgs and gauge sectors, while the SM fermions couple to the other one. Under these conditions, maximal entanglement is recovered when the $\mathrm{Z}_2$ symmetry holds exactly, implying equal masses between the fermion, gauge boson and Higgs boson in both sectors, respectively; the gauge, quartic, and Yukawa couplings of the two sectors also need to be equal, respectively.  This leads to the existence of a whole new mirror sector with identical properties to the SM ones.

It is worth commenting that we did not consider turning on gauge interactions in the case of the  electroweak symmetric phase because the enhanced global symmetry $U(2)\times U(2)$ is only approximate and in general violated by the mass terms in the scalar potential. However, from our discussion of adding gauge interactions in the broken phase, it should be clear that a smooth limit of sending the VEV to zero can be achieved from the perspective of entanglement maximization. That is, starting in the broken phase with $U(2)\times U(2)$ global symmetry and upon gauging the $SU(2)\times SU(2)$ subgroup, one could return to the unbroken phase by dialing the VEV $v\to 0$ while still maintaining maximal entanglement. In this case the 2-to-2 scattering $\Phi_a\Phi_b\to \Phi_c\Phi_d$ contains contributions from the scalar quartic diagrams as well as $s/t/u$-channels mediated by the massless gauge bosons. The requirement of maximal entanglement can be satisfied with an emerging $\mathrm{Z}_2$ symmetry enforcing the equality of the two gauge coupling strengths.

Last but not least, it is intriguing that, starting with one $U(2)$ symmetry group in the general 2HDM, our procedure of entanglement maximization enlarges the symmetry to twice the amount: $U(2)\times U(2)$. This doubling of symmetry persists when we introduce gauge interactions, as long as a $\mathrm{Z}_2$ symmetry which interchanges the two gauge groups is introduced. In other words, entanglement maximization doubles the amount of symmetry, either local or global, that is initially present. The observation is reminiscent of the two-qubit repetition code employed in quantum error correction as discussed in Appendix \ref{app:repetition_code}. How maximal entanglement produces the repetition of symmetries in more generic settings will be investigated in future works.

\acknowledgements

This work is supported in part by the U.S.
Department of Energy, Office of High Energy Physics, under contract DE-AC02-06CH11357 at Argonne, and under the Quantum Information Science Enabled Discovery (QuantISED)  program contract No.~89243024CSC000002. 
M.C. research at Perimeter Institute is supported in part by the Government of Canada through the Department of Innovation, Science and Economic Development, and by the Province of Ontario through the Ministry of Colleges and Universities. I.L. is  supported in part by the U.S. Department of Energy, Office of Nuclear
Physics, under grant DE-SC0023522 at Northwestern. The work of C.W.\ at the University of Chicago has been supported by the DOE grant DE-SC0013642. W.L. is supported by the Department of Energy through the Fermilab QuantiSED program in the area of ``Intersections of QIS and Theoretical Particle Physics". C.W. and M.C. would like to thank the Aspen Center for Physics, which is supported by National Science Foundation grant No.~PHY-1607611, where part of this work has been done. The work of G.C.
is supported by the Swiss National Science Foundation (No. PP00P21 76884).
\appendix
\renewcommand{\thesubsection}{\Alph{subsection}}
\renewcommand{\thesubsubsection}{\alph{subsubsection}}

\section{Scattering Amplitudes in Different Channels}
\label{sect:app}

We present the scattering amplitudes in the broken phase for the remaining initial states omitted from the main text. These results correspond to the symmetry configurations discussed in this work, specifically $Z_i = \pm Z_5$. Conventions follow Eq.~(\ref{eq:amp_broken}).

\tocless\subsubsection{$H^+ \tilde{H}^0 \to H^+ \tilde{H}^0$}
\begin{align}
   {\cal M}^{(2)}_0 &= Z
   \begin{bmatrix}
    1 & 0 & 0 & 1 \\[-.3cm]
    0 & 1 & \pm 1 & 0 \\[-.3cm]
    0 & \pm 1 & 1 & 0 \\[-.3cm]
    1 & 0 & 0 & 1 
   \end{bmatrix} \ ; & & 
   \\
   {\cal M}^{(2)}_{s, G^\pm}&= Z^2
   \begin{bmatrix}
    1 & 0 & 0 & 1 \\[-.3cm]
    0 & 0 & 0 & 0 \\[-.3cm]
    0 & 0 & 0 & 0 \\[-.3cm]
    1 & 0 & 0 & 1 \\
   \end{bmatrix}\ ,
   &
   {\cal M}^{(2)}_{s, H^\pm} &=
   Z^2 \begin{bmatrix}
    0 & 0 & 0 & 0 \\[-.3cm]
    0 & 1 & \pm 1 & 0 \\[-.3cm]
    0 & \pm 1 & 1 & 0 \\[-.3cm]
    0 & 0 & 0 & 0 \\
   \end{bmatrix} \ ;
   \\
   {\cal M}^{(2)}_{u,G^\pm} &=
   Z^2 \begin{bmatrix}
    1 & 0 & 0 & 0 \\[-.3cm]
    0 & 0 & \pm 1 & 0 \\[-.3cm]
    0 & \pm 1 & 0 & 0 \\[-.3cm]
    0 & 0 & 0 & 1 \\
   \end{bmatrix}\ ,
   &
   {\cal M}^{(2)}_{u,H^\pm} &=
   Z^2 
   \begin{bmatrix}
    0 & 0 & 0 & 1 \\[-.3cm]
    0 & 1 & 0 & 0 \\[-.3cm]
    0 & 0 &1 & 0 \\[-.3cm]
   1 & 0 & 0 & 0 \\
   \end{bmatrix}\ ;
   \\
   {\cal M}^{(2)}_{t,h} &= Z^2
   \begin{bmatrix}
    1 & 0 & 0 & 0 \\[-.3cm]
    0 & 1 & 0 & 0 \\[-.3cm]
    0 & 0 & 1 & 0 \\[-.3cm]
    0 & 0 & 0 & 1 \\
   \end{bmatrix}\ ,
   &
   {\cal M}^{(2)}_{t,G^0} &= 
   \begin{bmatrix}
    0 & 0 & 0 & 0 \\[-.3cm]
    0 & 0 & 0 & 0 \\[-.3cm]
    0 & 0 & 0 & 0 \\[-.3cm]
    0 & 0 & 0 & 0 \\
   \end{bmatrix} \ ,
   \\ \nonumber 
   {\cal M}^{(2)}_{t,H/A} &= Z^2
   \begin{bmatrix}
    0 & 0 & 0 & 1\\[-.3cm]
    0 & 0 & \pm 1 & 0 \\[-.3cm]
    0 & \pm 1 & 0 & 0 \\[-.3cm]
    1 & 0 & 0 & 0 \\
   \end{bmatrix}\ ,
  &
  {\cal M}^{(2)}_{t,A/H} &=
   \begin{bmatrix}
    0 & 0 & 0 & 0 \\[-.3cm]
    0 & 0 & 0 & 0 \\[-.3cm]
    0 & 0 & 0 & 0 \\[-.3cm]
    0 & 0 & 0 & 0 \\
   \end{bmatrix} \ .
\end{align}

\tocless\subsubsection{$H^+ H^- \to H^+ H^-$}

\begin{align}
   {\cal M}^{(3)}_0 &= 2 \, Z
   \begin{bmatrix}
    1 & 0 & 0 & 1 \\[-.3cm]
    0 & 1 & \pm 1 & 0 \\[-.3cm]
    0 & \pm 1 & 1 & 0 \\[-.3cm]
    1 & 0 & 0 & 1 
   \end{bmatrix} \ ; & & 
   \\
   {\cal M}^{(3)}_{s, h}&= \frac{Z^2}{2}
   \begin{bmatrix}
    1 & 0 & 0 & 1 \\[-.3cm]
    0 & 0 & 0 & 0 \\[-.3cm]
    0 & 0 & 0 & 0 \\[-.3cm]
    1& 0 & 0 & 1 \\
   \end{bmatrix}\ ,
   &
   {\cal M}^{(3)}_{s, G^0} &=
   \begin{bmatrix}
    0 & 0 & 0 & 0 \\[-.3cm]
    0 & 0 & 0 & 0 \\[-.3cm]
    0 & 0 & 0 & 0 \\[-.3cm]
    0 & 0 & 0 & 0 \\
   \end{bmatrix} \ ,
   \\ \nonumber
   {\cal M}^{(3)}_{s, H/A} &= \frac{Z^2}{2} \begin{bmatrix}
    0 & 0 & 0 & 0 \\[-.3cm]
    0 & 1 & \pm 1 & 0 \\[-.3cm]
    0 & \pm 1 & 1 & 0 \\[-.3cm]
    0 & 0 & 0 & 0 \\
   \end{bmatrix}\ ,
   &
   {\cal M}^{(3)}_{s, A/H} &=
   \begin{bmatrix}
    0 & 0 & 0 & 0 \\[-.3cm]
    0 & 0 & 0 & 0 \\[-.3cm]
    0 & 0 & 0 & 0 \\[-.3cm]
   0 & 0 & 0 & 0 \\
   \end{bmatrix}\ ;
   \\
   {\cal M}^{(3)}_{t,h} &= \frac{Z^2}{2}
   \begin{bmatrix}
    1 & 0 & 0 & 0 \\[-.3cm]
    0 & 1 & 0 & 0 \\[-.3cm]
    0 & 0 & 1 & 0 \\[-.3cm]
    0 & 0 & 0 & 1 \\
   \end{bmatrix}\ ,
   &
   {\cal M}^{(3)}_{t,G^0} &= 
   \begin{bmatrix}
    0 & 0 & 0 & 0 \\[-.3cm]
    0 & 0 & 0 & 0 \\[-.3cm]
    0 & 0 & 0 & 0 \\[-.3cm]
    0 & 0 & 0 & 0 \\
   \end{bmatrix} \ , 
   \\ \nonumber 
   {\cal M}^{(3)}_{t,H/A} &= \frac{Z^2}{2}
   \begin{bmatrix}
    0 & 0 & 0 & 1\\[-.3cm]
    0 & 0 & \pm 1 & 0 \\[-.3cm]
    0 & \pm 1 & 0 & 0 \\[-.3cm]
   1 & 0 & 0 & 0 \\
   \end{bmatrix}\ ,
  &
  {\cal M}^{(3)}_{t,A/H} &=
   \begin{bmatrix}
    0 & 0 & 0 & 0 \\[-.3cm]
    0 & 0 & 0 & 0 \\[-.3cm]
    0 & 0 & 0 & 0 \\[-.3cm]
    0 & 0 & 0 & 0 \\
   \end{bmatrix} \ . 
\end{align}

\tocless\subsubsection{$H^0 \tilde{H}^0 \to H^0 \tilde{H}^0$}

\begin{align}
   {\cal M}^{(4)}_0 &= 2 \, Z
   \begin{bmatrix}
    1 & 0 & 0 & 1 \\[-.3cm]
    0 & 1 & \pm 1 & 0 \\[-.3cm]
    0 & \pm 1 & 1 & 0 \\[-.3cm]
    1 & 0 & 0 & 1 
   \end{bmatrix} \ ; & & 
   \\
   {\cal M}^{(4)}_{s, h}&= 2 \, Z^2
   \begin{bmatrix}
    1 & 0 & 0 & 1 \\[-.3cm]
    0 & 0 & 0 & 0 \\[-.3cm]
    0 & 0 & 0 & 0 \\[-.3cm]
    1& 0 & 0 & 1 \\
   \end{bmatrix}\ ,
   &
   {\cal M}^{(4)}_{s, G^0} &=
   \begin{bmatrix}
    0 & 0 & 0 & 0 \\[-.3cm]
    0 & 0 & 0 & 0 \\[-.3cm]
    0 & 0 & 0 & 0 \\[-.3cm]
    0 & 0 & 0 & 0 \\
   \end{bmatrix} \ ,
   \\ \nonumber
   {\cal M}^{(4)}_{s, H/A} &= 2 \, Z^2 \begin{bmatrix}
    0 & 0 & 0 & 0 \\[-.3cm]
    0 & 1 & \pm 1 & 0 \\[-.3cm]
    0 & \pm 1 & 1 & 0 \\[-.3cm]
    0 & 0 & 0 & 0 \\
   \end{bmatrix}\ ,
   &
   {\cal M}^{(4)}_{s, A/H} &=
   \begin{bmatrix}
    0 & 0 & 0 & 0 \\[-.3cm]
    0 & 0 & 0 & 0 \\[-.3cm]
    0 & 0 & 0 & 0 \\[-.3cm]
   0 & 0 & 0 & 0 \\
   \end{bmatrix}\ ;
   \\
   {\cal M}^{(4)}_{u,h/G^0} &= 2 \, Z^2
   \begin{bmatrix}
    1 & 0 & 0 & 0 \\[-.3cm]
    0 & 0 & \pm 1 & 0 \\[-.3cm]
    0 & \pm 1 & 0 & 0 \\[-.3cm]
    0 & 0 & 0 & 1 \\
   \end{bmatrix}\ ,
   &
   {\cal M}^{(4)}_{u,H/A} &= 2 \, Z^2
   \begin{bmatrix}
    0 & 0 & 0 & 1 \\[-.3cm]
    0 & 1 & 0 & 0 \\[-.3cm]
    0 & 0 & 1 & 0 \\[-.3cm]
    1 & 0 & 0 & 0 \\
   \end{bmatrix} \ ;
   \\
   {\cal M}^{(4)}_{t,h} &= 2 \, Z^2
   \begin{bmatrix}
    1 & 0 & 0 & 0\\[-.3cm]
    0 & 1 & 0 & 0 \\[-.3cm]
    0 & 0 & 1 & 0 \\[-.3cm]
    0 & 0 & 0 & 1 \\
   \end{bmatrix}\ ,
  &
  {\cal M}^{(4)}_{t,G^0} &=
   \begin{bmatrix}
    0 & 0 & 0 & 0 \\[-.3cm]
    0 & 0 & 0 & 0 \\[-.3cm]
    0 & 0 & 0 & 0 \\[-.3cm]
    0 & 0 & 0 & 0 \\
   \end{bmatrix} \ ,
   \\ \nonumber
   {\cal M}^{(4)}_{t,H/A} &= 2 \, Z^2
   \begin{bmatrix}
    0 & 0 & 0 & 1\\[-.3cm]
    0 & 0 & \pm 1 & 0 \\[-.3cm]
    0 & \pm 1 & 0 & 0 \\[-.3cm]
    1 & 0 & 0 & 0 \\
   \end{bmatrix}\ ,
  &
  {\cal M}^{(4)}_{t,A/H} &=
   \begin{bmatrix}
    0 & 0 & 0 & 0 \\[-.3cm]
    0 & 0 & 0 & 0 \\[-.3cm]
    0 & 0 & 0 & 0 \\[-.3cm]
    0 & 0 & 0 & 0 \\
   \end{bmatrix} \ . 
\end{align}

\tocless\subsubsection{$H^0 \tilde{H}^0 \to H^+ H^- \; / \; H^+ H^- \to H^0 \tilde{H}^0$}

\begin{align}
   {\cal M}^{(5)}_0 &= Z
   \begin{bmatrix}
    1 & 0 & 0 & 1 \\[-.3cm]
    0 & 1 & \pm 1 & 0 \\[-.3cm]
    0 & \pm 1 & 1 & 0 \\[-.3cm]
    1 & 0 & 0 & 1 
   \end{bmatrix} \ ; & & 
   \\
   {\cal M}^{(5)}_{s, h}&= Z^2
   \begin{bmatrix}
    1 & 0 & 0 & 1 \\[-.3cm]
    0 & 0 & 0 & 0 \\[-.3cm]
    0 & 0 & 0 & 0 \\[-.3cm]
    1& 0 & 0 & 1 \\
   \end{bmatrix}\ ,
   &
   {\cal M}^{(5)}_{s, G^0} &=
   \begin{bmatrix}
    0 & 0 & 0 & 0 \\[-.3cm]
    0 & 0 & 0 & 0 \\[-.3cm]
    0 & 0 & 0 & 0 \\[-.3cm]
    0 & 0 & 0 & 0 \\
   \end{bmatrix} \ ,
   \\ \nonumber
   {\cal M}^{(5)}_{s, H/A} &= Z^2 \begin{bmatrix}
    0 & 0 & 0 & 0 \\[-.3cm]
    0 & 1 & \pm 1 & 0 \\[-.3cm]
    0 & \pm 1 & 1 & 0 \\[-.3cm]
    0 & 0 & 0 & 0 \\
   \end{bmatrix}\ ,
   &
   {\cal M}^{(5)}_{s, A/H} &=
   \begin{bmatrix}
    0 & 0 & 0 & 0 \\[-.3cm]
    0 & 0 & 0 & 0 \\[-.3cm]
    0 & 0 & 0 & 0 \\[-.3cm]
   0 & 0 & 0 & 0 \\
   \end{bmatrix}\ ;
   \\
   {\cal M}^{(5)}_{u,G^\pm} &= Z^2
   \begin{bmatrix}
    1 & 0 & 0 & 0 \\[-.3cm]
    0 & 0 & \pm 1 & 0 \\[-.3cm]
    0 & \pm 1 & 0 & 0 \\[-.3cm]
    0 & 0 & 0 & 1 \\
   \end{bmatrix}\ ,
   &
   {\cal M}^{(5)}_{u,H^\pm} &= Z^2
   \begin{bmatrix}
    0 & 0 & 0 & 1 \\[-.3cm]
    0 & 1 & 0 & 0 \\[-.3cm]
    0 & 0 & 1 & 0 \\[-.3cm]
    1 & 0 & 0 & 0 \\
   \end{bmatrix} \ ;
   \\
   {\cal M}^{(5)}_{t,G^\pm} &= Z^2
   \begin{bmatrix}
    1 & 0 & 0 & 0 \\[-.3cm]
    0 & 1 & 0 & 0 \\[-.3cm]
    0 & 0 & 1 & 0 \\[-.3cm]
    0 & 0 & 0 & 1 \\
   \end{bmatrix}\ ,
  &
  {\cal M}^{(5)}_{t,H^\pm} &= Z^2
   \begin{bmatrix}
    0 & 0 & 0 & 1 \\[-.3cm]
    0 & 0 & \pm 1 & 0 \\[-.3cm]
    0 & \pm 1 & 0 & 0 \\[-.3cm]
    1 & 0 & 0 & 0 \\
   \end{bmatrix} \ .
\end{align}
\section{Quantum Repetition Code and Strong $\mathrm Z_2$ Symmetry}\label{app:repetition_code}
In this appendix, we discuss the quantum information properties of the maximally entangling matrices. We will focus on Eq.~(\ref{eq:XX1}) as an example as the other differ only by a basis transformation. One can easily check that the output of the channel is always $a \ket{++}+b\ket{--}$, regardless of the initial state. This is equivalent to    the 2-qubit repetition code\cite{eczoo_quantum_repetition,knill2000efficientlinearopticsquantum}, which we will review briefly here.

An n-qubit repetition code uses $n$ physical qubits to encode the information of one qubit, by only allowing $a\ket{0\dots 0}+b\ket{1\dots 1}$ as the ``good states". In other words, only the 2-dimensional subspace $span\{\ket{0\dots 0},\ket{1\dots1}\}$ is utilized, which is called the code space. All states in the code space are invariant under the $Z_2$ symmetries generated by the operators $\sigma_z^{(i)}\sigma_z^{(j)}$ on any two qubits $i,j$. By measuring $\sigma_z^{(i)}\sigma_z^{(j)}$, one can one can detect bit-flip errors $\sigma_x$ on individual physical qubits. If bit-flip errors occur on fewer than half of the physical qubits, one can correct for such errors. Repetition can also be in the eigenbasis of $\sigma_1$ by using a different code space $span\{\ket{+\dots +},\ket{-\dots -}\}$. The states in the code space are invariant under $\sigma_x^{(i)}\sigma_x^{(j)}$, and measuring such product operators allows the detection and correction of phase-flip errors $\sigma_z$.

\begin{figure}[ht]
    \centering
    \includegraphics[width=0.5\linewidth]{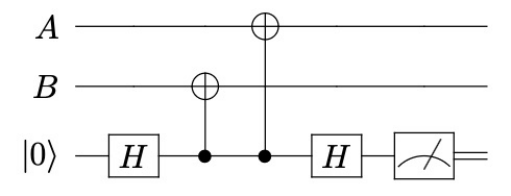}
    \caption{The quantum circuit to detect phase-flip errors in a 2-qubit repetition code}
    \label{fig:QCircuit}
\end{figure}

With a 2-qubit repetition code in the $\sigma_1$ basis, a phase-flip error  on one physical qubit can be detected but cannot be corrected, as we cannot know whether the problematic state $\ket{+-}$ should be corrected to $\ket{++}$ or $\ket{--}$. The quantum channel to detect the error is \fig{QCircuit}.
The qubits $A,B$ are the data qubits that carry the information, and the third is an auxiliary qubit, which is introduced to allow the measurement of $\sigma_1^A\sigma_1^B$ without measuring each qubit directly. Denoting the initial state of $A$ and $B$ as $\psi_{AB}$, after the application of the first Hadamard gate
\begin{equation}
    H=\frac{1}{\sqrt{2}}\left [\begin{array}{cc}
    1 & 1  \\
    1 & -1
\end{array}\right],
\end{equation}
and the CNOT gates
\begin{equation}
    \mathrm{CNOT}=\left [\begin{array}{cc}
    1 & 0  \\
    0 & 0
\end{array}\right]_{\mathrm{control}}\otimes \mathbb I_{\mathrm{target}}+\left [\begin{array}{cc}
    0 & 0  \\
    0 & 1
\end{array}\right]_{\mathrm{control}}\otimes{[\sigma_1]}_{\mathrm{target}},
\end{equation}
the state of the three qubit is \begin{equation}
\frac 1 {\sqrt{2}}\ket{\psi_{AB}}\otimes \ket{0}+\frac 1 {\sqrt{2}}(\sigma_1^A \otimes \sigma_1^B\ket{\psi_{AB}})\otimes \ket{1}.
\end{equation}
After the second Hadamard gate, the state of the three qubit becomes 
\begin{equation}
\frac 1 2(\mathbb I+ \sigma_1^A \otimes\sigma_1^B)\ket{\psi_{AB}}\otimes \ket{0}+\frac 1 2(\mathbb I- \sigma_1^A \otimes\sigma_1^B)\ket{\psi_{AB}}\otimes \ket{1}.
\end{equation}
Afterwards, one measures the ancillary qubit to detect whether the 2-qubit system is in the code space. If the result is 0, the $AB$ system is projected to the code space; if the result is 1, the state is projected into the space orthogonal to the code space and should be discarded. This process is the post-selection.

We notice that the channel Eq.~(\ref{eq:XX1}) is equivalent to the combination of error detection via \fig{QCircuit} and post-selection.  It keeps the part of the wavefunction that is invariant under the $Z_2$ group generated by $\tau^1\otimes\tau^1$, and discards the part that is not invariant. In other words, the channel is only supported in such a $Z_2$ symmetric subspace. 

This observation helps us to clarify what symmetry can enforce the maximal entanglement among the computational basis states. To do this, we also need the notion of \textit{strong symmetries}. Ordinarily, symmetries act on operators via conjugation $O\to UOU^\dagger$ , where $U$ is a unitary transformation. Using the notion in recent works \cite{PRXQuantum.6.010344}, the operator $O$ has strong symmetry if it is further invariant under left or right multiplication alone up to a phase factor, i.e. $UO=e^{i\theta}O$ and $OU^\dagger=e^{-i\theta}O$. This implies that $O$ is supported in a subspace that transforms with the same phase under $U$. 

We observe that $\mathcal{M}_0^{I}$, $\mathcal{M}_0^{3/4/5}$, and $\mathcal{M}_0^{1/2}$ all have strong $Z_2$ symmetries that are generated by $[\tau^1\otimes\tau^1]$ or $[\tau^2\otimes\tau^2]$:
\begin{align}
    &[\tau^1\otimes\tau^1] \mathcal{M}_0^{I}=\mathcal{M}_0^{I}=\mathcal{M}_0^{I}[\tau^1\otimes\tau^1 ]\\
     & [\tau^2\otimes\tau^2] \mathcal{M}_0^{1/2}=\mathcal{M}_0^{1/2}=\mathcal{M}_0^{1/2}[\tau^2\otimes\tau^2 ]\\
    & [\tau^2\otimes\tau^2] \mathcal{M}_0^{3/4/5}=-\mathcal{M}_0^{3/4/5}=\mathcal{M}_0^{3/4/5}[\tau^2\otimes\tau^2 ]
\end{align}
To see what we gain with strong symmetries, observe that any linear combination  $\alpha\mathcal{M}_0^{3/4/5}+\beta\mathcal{M}_0^{1/2}$ satisfies the ordinary (weak) symmetry by conjugation, whereas it does not satisfy the strong symmetry due to the different signs, unless only one term is present. Thus, the requirement of strong symmetry is essential to enforce maximal entanglement. 

Therefore, we can conclude that along with the $U(2)\times U(2)$ symmetry of the Lagrangian, the strong $Z_2$ symmetry of the $T$-matrix generated by $[\tau^{1,2}\otimes\tau^{1,2}]$ enforces maximal entanglement among computational basis states. 
\section{Entanglement Among Generic Initial Product States}\label{app:gen_ent}

It is illuminating to consider the entanglement generated by Eqs.~(\ref{eq:mm1}) and (\ref{eq:mm2}) for general initial product states. Specifically, consider the following parameterization of a general product state,
\be
|\psi\rangle=\left(\cos({\theta_1}/2)\,|0\rangle_A +\sin({\theta_1}/2) \ e^{i\phi_1}\,|1\rangle_A \right)\otimes\left(\cos({\theta_2}/2)\,|0\rangle_B +\sin({\theta_2}/2) \ e^{i\phi_2}\,|1\rangle_B \right)\ ,
\ee
where we have used the Block sphere representation for both qubit-A and qubit-B. We can then proceed to compute the concurrence of the outgoing states and demand $\Delta({\cal M}|\psi\rangle)=1$, which leads to the following conditions,
\bea
\lambda_i=\lambda_5&:& \ \ \sin\theta_1 \cos\phi_1 + \sin\theta_2\cos\phi_2 = 0 \ , \\
\lambda_i=-\lambda_5&:& \ \ \sin\theta_1\, \sin\phi_1 + \sin\theta_2\sin\phi_2 = 0 \ .
\eea
These conditions are satisfied for $\theta_i=0, \pi$, which correspond to the four basis states in the computational basis. Moreover, they are fulfilled also when $\phi_i=\pm\pi/2$ for $\lambda_i=\lambda_5$ and $\phi_i=0$ for $\lambda_i=-\lambda_5$. In other words, by requiring the scattering process to maximize entanglement for the full computational basis, the resulting scattering amplitudes will also maximize entanglement for the following general product states:
\bea
|\psi+\rangle &=& \left(a|0\rangle_A +i\,b|1\rangle_A \right)\otimes\left(c|0\rangle_B +i\,d|1\rangle_B \right)\ ,\\
|\psi-\rangle &=& \left(a|0\rangle_A + b|1\rangle_A \right)\otimes\left(c|0\rangle_B + d|1\rangle_B \right)\ ,
\label{eq:imqubit}
\eea
where $\{a,b,c,d\}$ are arbitrary real coefficients and $|\psi\pm\rangle$ corresponds to the two possible solutions $\lambda_i=\pm \lambda_5$.


\bibliography{refs}

\end{document}